\documentclass[10pt,twocolumn,journal]{IEEEtran}
\usepackage{times}
\usepackage{amsbsy}
\usepackage{latexsym}
\usepackage{color}
\usepackage{amssymb}
\usepackage{amsmath}
\usepackage[ansinew]{inputenc}
\usepackage[dvips ]{graphicx}
\input epsf
\usepackage{epic,eepic}
\usepackage{url}

\title{\Huge Blind Adaptive Interference Suppression Based on
Set-Membership Constrained Constant-Modulus Algorithms with
Time-Varying Bounds }
\author{Rodrigo C. de Lamare and Paulo S. R. Diniz
\thanks{ Part of this work has been presented
at ICASSP 2006. This work was supported in part by CNPQ and FAPERJ.
Dr. Rodrigo C. de Lamare is with the Communications Research Group,
University of York, United Kingdom. Prof.  Paulo S. R. Diniz is with
LPS/COPPE-UFRJ, Rio de Janeiro, Brazil.  Emails: rcdl500@york.ac.uk,
diniz@lps.ufrj.br}}

\begin{document}
\maketitle
\begin{abstract}
This work presents blind constrained constant modulus (CCM) adaptive
algorithms based on the set-membership filtering (SMF) concept and
incorporates dynamic bounds {  for interference suppression}
applications. We develop stochastic gradient and recursive least
squares type algorithms based on the CCM design criterion in
accordance with the specifications of the SMF concept. We also
propose a blind framework that includes channel and amplitude
estimators that take into account parameter estimation dependency,
multiple access interference (MAI) and inter-symbol interference
(ISI) to address the important issue of bound specification in
multiuser communications. A convergence and tracking analysis of the
proposed algorithms is carried out along with the development of
analytical expressions to predict their performance. Simulations for
a number of scenarios of interest with a DS-CDMA system show that
the proposed algorithms outperform previously reported techniques
with a smaller number of parameter updates and a reduced risk of
overbounding or underbounding.

\end{abstract}


\section{Introduction}

Set-membership filtering (SMF)
\cite{huang,werner1,gollamudi,nagaraj} is a class of recursive
estimation algorithms that, on the basis of a pre-determined error
bound, seeks a set of parameters that yield bounded filter output
errors. These algorithms have been applied to a variety of
applications including adaptive equalization \cite{gollamudi} and
multi-access interference suppression \cite{nagaraj,gollamudi2}.
The SMF algorithms are able to combat conflicting requirements
such as fast convergence and low misadjustment by introducing a
modification on the objective function. These
algorithms exhibit reduced complexity due to data-selective
updates, which involve two steps: a) information evaluation and b)
update of parameter estimates. If the filter update does not occur
frequently and the information evaluation does not involve much
computational complexity, the overall complexity can be
significantly reduced.

Adaptive SMF algorithms usually achieve good convergence and
tracking performance due to the use of an adaptive step size or an
adaptive forgetting factor for each update. This translates into
reduced complexity due to the data selective updating. However, the
performance of SMF techniques depends on the error-bound
specification, which is very difficult to obtain in practice due to
the lack of knowledge of the environment and its dynamics. In
wireless networks characterized by non-stationary environments,
where users often enter and exit the system, it is very difficult to
choose an error bound and the risk of overbounding (when the error
bound is larger than the actual one) and underbounding (when the
error bound is smaller than the actual one) is significantly
increased, leading to a performance degradation. In addition, when
the measured noise in the system is time-varying and the multiple
access interference (MAI) and the intersymbol interference (ISI)
encountered by a receiver in a wireless network are highly dynamic,
the selection of an error-bound is further complicated. This is
especially relevant for low-complexity estimation problems
encountered in applications that include mobile units and wireless
sensor networks \cite{akyldiz,wang}, where the sensors have limited
signal processing capabilities and power consumption is of central
importance. These problems suggest the deployment of mechanisms to
automatically adjust the error bound in order to guarantee good
performance and a low update rate (UR).

\subsection{Prior and Related Work}

In this context, blind methods are appealing because they can
alleviate the need for training sequences or pilots, thereby
increasing the throughput and efficiency of wireless networks. In
particular, blind algorithms based on constrained optimization
techniques are important in several areas of signal processing and
communications such as beamforming and interference suppression. The
constrained optimization required in these applications deals with
linear constraints that correspond to prior knowledge of certain
parameters such as the direction of arrival (DoA) of user signals in
antenna array processing \cite{liberti} and the signature sequence
of the desired signal in DS-CDMA systems \cite{honig,jidf}.
Therefore, linear signal models and constraints can be used to
describe various wireless communications systems including
multi-input multi-output (MIMO) and orthogonal frequency-division
multiplexing (OFDM) systems. For instance, linear constraints that
incorporate the knowledge of user signatures of a DS-CDMA system can
also be used to exploit the knowledge of the spatial signatures of
MIMO systems. A number of blind algorithms with different trade-offs
between performance and complexity have been reported in the last
decades \cite{honig}-\cite{delamareccmmimo}.  {Local scattering,
synchronization and estimation errors, imperfectly calibrated arrays
and imprecisely known wave field propagation conditions are typical
sources of uncertainties that can lead to a performance degradation
of blind algorithms. The literature indicates that the CCM-based
algorithms \cite{miguez}-\cite{delamareccmmimo} have a superior
performance to algorithms based on the constrained minimum variance
(CMV) criterion \cite{honig}- \cite{delamaretsp}. The CCM-based
algorithms exploit prior knowledge about the constant modulus
property of signals like M-PSK, which results in an improved
performance over CMV-type techniques and a performance that is very
close to the linear minimum mean-square error (MMSE) training-based
techniques. Moreover, the CCM-type algorithms are robust against
errors in the effective signature sequence required for blind
parameter estimation, which prevents a severe performance
degradation in the presence of uncertainties. These features make
CCM-type algorithms excellent candidates for interference
suppression in wireless networks.} The need for the adjustment of
parameters (step size, forgetting factor) and the computational
complexity of CCM-based techniques calls for approaches like SMF.
Prior work on SMF blind algorithms for interference suppression is
very limited \cite{gollamudi2,gollamudi3}. The use of time-varying
bounds is also restricted to applications where one assumes that the
"true" error bound is constant \cite{deller} and to the
parameter-dependent error bound recently proposed in
\cite{guo,bhotto}. The time-varying bound techniques so far reported
are not blind and do not exploit these mechanisms for channel and
parameter estimation.

\subsection{Contributions}

In this work, we propose set-membership blind adaptive constrained
algorithms based on the CCM criterion. Preliminary results have been
reported in \cite{delamare_smccm}. In particular, we derive
stochastic gradient (SG) and recursive least squares (RLS)-type CCM
algorithms designed in accordance with the specifications of the SMF
concept. The second contribution is a low-complexity blind framework
for parameter estimation, and tracking of parameter evolution, MAI
and ISI that relies on simple estimation techniques and employs the
proposed set-membership CCM algorithms with a time-varying bound.
The third contribution of this work is an analysis of the
optimization problem that gives rise to the proposed algorithms
along with a mean-squared error (MSE) convergence and tracking
analysis using the energy conservation approach \cite{yousef} for
predicting the performance of the algorithms. A simulation study
considers an interference suppression application to DS-CDMA
systems, which compares the performance of the proposed and existing
algorithms, and discusses the main features of the algorithms.

This paper is structured as follows. Section II describes the
DS-CDMA system model and briefly reviews linearly constrained
receivers. Section III introduces the SM blind algorithms with
time-varying bound. Section IV proposes blind parameter dependent
and interference dependent bounds along with blind channel and
amplitude estimation algorithms. Section V is dedicated to the
analysis of the proposed algorithms. Section VI is devoted to the
presentation and discussion of numerical results, while Section VII
gives the conclusions.

\section{DS-CDMA system model and Linearly Constrained Receivers}

Let us consider the uplink of a symbol synchronous DS-CDMA system
with $K$ users, $N$ chips per symbol and $L_{p}$ propagation
paths. A synchronous model is assumed for simplicity since it
captures most of the features of more realistic asynchronous
models with small to moderate delay spreads. The modulation is
assumed to have constant modulus. Let us assume that the signal
has been demodulated at the base station, the channel is constant
during each symbol and the receiver is perfectly synchronized with
the main channel path. The received signal after filtering by a
chip-pulse matched filter and sampled at chip rate yields an
$M$-dimensional received vector at time $i$
\begin{equation}
\begin{split}
{\boldsymbol r}[i] & = \sum_{k=1}^{K} A_{k}[i]  {b}_{k}[i]
{\boldsymbol C}_{k} {\boldsymbol h}_k[i]  +
 {\boldsymbol{\eta}}_k[i] + {\boldsymbol n}[i], \label{recsignal}
\end{split}
\end{equation}
where $M=N+L_{p}-1$, ${\boldsymbol n}[i] = [n_{1}[i]
~\ldots~n_{M}[i]]^{T}$ is the complex Gaussian noise vector with
zero mean and $E[{\boldsymbol n}[i]{\boldsymbol n}^{H}[i]] =
\sigma^{2}{\boldsymbol I}$ independent and identically distributed
samples, $(.)^{T}$ and $(.)^{H}$ denote transpose and Hermitian
transpose, respectively, and $E[.]$ stands for expected value. The
user symbols are denoted by ${b}_{k}[i]$, the amplitude of user $k$
is $A_{k}$[i], and ${\boldsymbol{\eta}}_k[i]$ is the intersymbol
interference (ISI) for user $k$. The signature of user $k$ is
represented by ${\boldsymbol c}_{k} = [c_{k}(1) \ldots
c_{k}(N)]^{T}$, the $M\times L_{p}$ constraint matrix ${\boldsymbol
C}_{k}$ that contains one-chip shifted versions of the signature
sequence for user $k$ and the $L_{p}\times 1$ vector ${\boldsymbol
h}_k[i]$ with the multipath components are described by
\begin{equation}
{\boldsymbol C}_{k} = \left[\begin{array}{c c c }
c_{k}(1) &  & {\boldsymbol 0} \\
\vdots & \ddots & c_{k}(1)  \\
c_{k}(N) &  & \vdots \\
{\boldsymbol 0} & \ddots & c_{k}(N)  \\
 \end{array}\right],
 {\boldsymbol h}_k[i]=\left[\begin{array}{c} {h}_{k,0}[i]
\\ \vdots \\ {h}_{k,L_{p}-1}[i]\\  \end{array}\right].
\end{equation}
The MAI comes from the non-orthogonality between the received
signature sequences. {  The ISI originates from the multipath
propagation effects of the channel, depends on the length of the
channel response and how it is related to the length of the chip
sequence. We define $L_s$ as the ISI span, i.e., the number of
symbols affected by the channel. For $L_{p}=1,~ L_{s}=1$ (no ISI),
for $1<L_{p}\leq N, L_{s}=3$, for $N <L_{p}\leq 2N, L_{s}=5$, and so
on.} At time instant $i$ we will have ISI coming not only from the
previous time instants but also from the next symbols. The linear
model in (\ref{recsignal}) can be used to represent other wireless
communications systems including multi-input multi-output (MIMO) and
orthogonal frequency-division multiplexing (OFDM) systems. For
example, the user signatures of a DS-CDMA system are equivalent to
the spatial signatures of a MIMO system.

In order to describe the design of linearly constrained receivers,
we consider the received vector ${\boldsymbol r}[i]$, the $M\times
L_{p}$ constraint matrix ${\boldsymbol C}_k$ that contains one-chip
shifted versions of the signature sequence for user $k$ and the
$L_{p}\times 1$ vector ${\boldsymbol h}_k[i] = [{h}_{k,0}[i] ~
\ldots ~ {h}_{k,L_p-1}[i]]^{T}$ with the multipath components to be
estimated. The CCM linear receiver design is equivalent to
determining an FIR filter ${\boldsymbol w}_{k}[i]$ with $M$
coefficients that provide an estimate of the desired symbol as
follows
\begin{equation}
z_k[i] = {\boldsymbol w}_{k}^{H}[i]{\boldsymbol r}[i],
\label{symbol}
\end{equation}
where the detected symbol is given by $\hat{b}_{k}[i] =
Q({\boldsymbol w}_{k}^{H}[i]{\boldsymbol r}[i])$, where ${\rm
Q}(\cdot)$ is a function that performs the detection according to
the constellation employed.

The design of the receive filter ${\boldsymbol w}_{k}[i]$ is based
on the optimization of the CM cost function
\begin{equation}
J_{CM}({\boldsymbol w}_{k}[i]) = E\Big[(|{\boldsymbol
w}_{k}^{H}[i]{\boldsymbol r}[i]|^{2}-1)^{2}\Big] \label{cost}
\end{equation}
subject to the constraints given by ${\boldsymbol
w}_{k}^{H}[i]{\boldsymbol p}_{k}[i] = \nu $, where ${\boldsymbol
p}_{k}[i] = {\boldsymbol C}_{k}{\boldsymbol h}_k[i] $ { is the
effective signature vector that corresponds to the convolution
between the original signature sequence and the channel gains}, and
$\nu$ is a constant to ensure the convexity of the optimization
problem as will be discussed later on. This approach assumes the
knowledge of the channel. However, when multipath is present these
parameters are unknown and time-varying, requiring channel
estimation. The CCM receive filter expression that iteratively
solves the constrained optimization problem in (\ref{cost}) is given
by
\begin{equation}
\begin{split}
{\boldsymbol w}_{k}[i+1] & = {\boldsymbol
R}^{-1}_{k}[i]\Big[{\boldsymbol d}_{k}[i]-\Big({\boldsymbol
p}_{k}^{H}[i]{\boldsymbol R}^{-1}_{k}[i]{\boldsymbol
p}_{k}[i]\Big)^{-1}
\\ &\quad \cdot \Big({\boldsymbol
p}_{k}^{H}[i]{\boldsymbol R}^{-1}_{k}[i] {\boldsymbol d}_{k}[i]
-\nu\Big){\boldsymbol p}_{k}[i]\Big], ~~~ i = 1,2, \ldots
\label{ccmexp}
\end{split}
\end{equation}
where $z_{k}[i]={\boldsymbol w}_{k}^{H}[i]{\boldsymbol r}[i]$,
${\boldsymbol R}_{k}[i]=E[|z_{k}[i]|^{2}{\boldsymbol
r}[i]{\boldsymbol r}^{H}[i]]$, ${\boldsymbol
d}_{k}[i]=E[z_{k}^{*}[i]{\boldsymbol r}[i]]$. A detailed
derivation of the CCM estimation approach can be found in
\cite{delamareccm,mswfccm,delamareccmmimo}. It should be remarked
that the expression in (\ref{ccmexp}) is a function of previous
values of the filter ${\boldsymbol w}_{k}[i]$ and therefore must
be iterated in order to reach a solution. In addition to this, the
iterative method in (\ref{ccmexp}) assumes the knowledge of the
channel parameters. Since there is a large number of applications
that have to deal with unknown multipath propagation, it is also
important to be able to blindly estimate the multipath components.

In order to blindly estimate the channel, a designer can adopt the
blind channel estimation procedure based on the subspace approach
reported in \cite{xutsa,douko} and which is described by
\begin{equation}
{\hat{\boldsymbol h}}_{k}[i] = \arg \min_{{\boldsymbol h}_k[i]} ~
{ {\boldsymbol h}_{k}^{H}[i] {\boldsymbol C}^{H}_{k}{\boldsymbol
R}^{-1}[i]{\boldsymbol C}_{k}{\boldsymbol h}_{k}[i] }
\end{equation}
subject to $||{\boldsymbol h}_k[i]||=1$, where ${\boldsymbol
R}[i]=E[{\boldsymbol r}[i]{\boldsymbol r}^{H}[i]]$. The solution is
the eigenvector of the $L_{p}\times L_{p}$ matrix corresponding to
the minimum eigenvalue of $ {\boldsymbol C}^{H}_{k}{\boldsymbol
R}^{-1}[i]{\boldsymbol C}_{k}$ obtained by an eigenvalue
decomposition (EVD). Here, we use ${\boldsymbol R}_{k}[i]$ in lieu
of ${\boldsymbol R}[i]$ to avoid the estimation of both
${\boldsymbol R}[i]$ and ${\boldsymbol R}_{k}[i]$, which shows no
performance loss as reported and investigated in
\cite{delamareccm,mswfccm,delamareccmmimo}.

\section{Set-Membership Blind Adaptive Constrained Algorithms with Time Varying Error Bounds}

In this section, we describe an adaptive filtering framework that
combines the set-membership (SM) concept with blind constrained
algorithms based on the CCM design. We also introduce simple
time-varying error bounds to take into consideration the evolution
of the receive filter and the MAI and ISI effects. In SM filtering
\cite{huang}, the parameter vector ${\boldsymbol w}_{k}[i]$ for
user $k$ is designed to achieve a specified bound on the magnitude
of an estimated quantity $z_{k}[i]$. As a result of this
constraint, the SM blind adaptive algorithm will only perform
filter updates for certain data. Let $\Theta [i]$ represent the
set containing all ${\boldsymbol w}_{k}[i]$ that yield an
estimation quantity upper bounded in magnitude by a time-varying
error bound $\gamma_k[i]$. Thus, we can write
\begin{equation}
\Theta[i] = \bigcap_{{({\boldsymbol r}[i])\in \mathcal{S}}} \{
{\boldsymbol w}_{k} \in {\mathcal{C}}^{M}:\mid z_{k}[i]\mid \leq
\gamma_k[i] \},
\end{equation}
where ${\boldsymbol r}[i]$ is the observation vector, $\mathcal{S}$
is the set of all possible data pairs $(b_{k}[i],~ {\boldsymbol
r}[i])$ and the set $\Theta[i]$ is referred to as the feasibility
set, and any point in it is a valid estimate $z_{k}[i]={\boldsymbol
w}_{k}^{H}[i]{\boldsymbol r}[i]$. Since it is not practical to
predict all data pairs, adaptive methods work with the membership
set $\psi_{i}= \bigcap_{m=1}^{i} {\mathcal{H}}_{m}$ provided by the
observations, where ${\mathcal{H}}_{m}=\{{\boldsymbol w}_{k} \in
{\mathcal{C}}^{M}: |z_{k}[i]| \leq \gamma_k[i]\}$. In order to
devise an effective SM algorithm, the bound $\gamma_k[i]$ must be
appropriately chosen. Due to the time-varying nature of many
practical environments, this bound should also be adaptive and
capable of estimating certain characteristics for the SM estimation
technique.  {We devise SM-CCM algorithms equipped with variable step
sizes and forgetting factors that are able to automatically tune to
different situations, which is an advantage over methods operating
with fixed parameters in time-varying scenarios.} In what follows,
we derive SM-CCM algorithms for blind parameter estimation that
assume time-varying error bounds.

\subsection{Set-Membership CCM Stochastic Gradient-Type Algorithm}

Here we develop the set-membership CCM stochastic gradient-type
(SM-CCM-SG) algorithm. The basic idea is to devise a gradient
descent strategy to compute a parameter vector ${\boldsymbol w}_{k}$
for user $k$ that minimizes the instantaneous CM cost function
subject to certain constraints, which requires for adaptation that
the square of the error $e_k^2[i]$ exceeds a specified error bound
$\gamma_k^2[i]$. Mathematically, the proposed SM-CCM-SG algorithm
solves the following optimization problem {
\begin{equation}
\begin{split}
{\rm minimize}~~~  J_{CM}({\boldsymbol w}_{k}[i])  & = (|{\boldsymbol w}_{k}^{H}[i]{\boldsymbol r}[i]|^{2}
-1)^{2} = e^2_k[i]  \\
{\rm subject}~~{\rm to} ~~~ {\boldsymbol w}_{k}^H[i] {\boldsymbol
p}_k[i] & = \nu \\  {\rm whenever}~~  e^2_k[i] & \geq
\gamma_k^{2}[i]
\end{split}
\end{equation}
This problem can be solved using the method of Lagrange multipliers
using the equality constraint $e^2_k[i] = \gamma_k^2[i]$
\cite{diniz}:
\begin{equation}
\begin{split}
{\mathcal L}({\boldsymbol w}_k[i],\kappa_k) & = (|{\boldsymbol
w}_k^H[i] {\boldsymbol r}[i]|^2 -1 )^2  +
[({\boldsymbol w}_k^H[i]{\boldsymbol p}_k[i] - \nu)\kappa_k^*] \\
& \quad + [ \kappa_k^*({\boldsymbol p}_k^H[i]{\boldsymbol w}_k[i] -
\nu)], 
\end{split} \label{lag1}
\end{equation}
where $\kappa_k$ is a Lagrange multiplier. Computing the gradient
terms of (\ref{lag1}) and equating them to zero, we obtain
\begin{equation}
\begin{split}
\nabla_{{\boldsymbol w}_k^*[i]}{\mathcal L}({\boldsymbol
w}_k[i],\kappa_k) & = 2(e_k[i] {\boldsymbol r}[i] z_k^*[i]) + {\boldsymbol p}_k[i] \kappa_k  = 0 \\
\nabla_{\kappa_k^*}{\mathcal L}({\boldsymbol w}_k[i],\kappa_k) & =
{\boldsymbol w}_k^H[i]{\boldsymbol
p}_k[i] - \nu = 0. 
\label{systeq}
\end{split}
\end{equation}
We employ a gradient descent rule to solve for the above
equations. Using the first equation of (\ref{systeq}), we have
\begin{equation}
{\boldsymbol w}_k[i+1] = {\boldsymbol w}_k[i] - \mu_k (e_k[i]
z_k^*[i] {\boldsymbol r}[i]  + {\boldsymbol p}_k[i] \kappa_k)
\end{equation}
Using the second equation of (\ref{systeq}), we obtain
\begin{equation}
{\boldsymbol w}_k[i+1] = {\boldsymbol \Pi}_k[i]({\boldsymbol
w}_k[i] - \mu_k e_k[i] z_k^*[i] {\boldsymbol r}[i]) + \nu
{\boldsymbol p}_k[i]({\boldsymbol p}_k^H[i] {\boldsymbol
p}_k[i])^{-1}, \label{gradrec}
\end{equation}
where $\mu_k$ is the effective step size, $e_k[i] = |z_k[i]|^2-1$ is
the error signal for user $k$, and ${\boldsymbol \Pi}_k[i] =
{\boldsymbol I} - {\boldsymbol p}_k[i]({\boldsymbol p}_k^H[i]
{\boldsymbol p}_k[i])^{-1} {\boldsymbol p}_k^H[i]$ is a projection
matrix that ensures the constraint and ${\boldsymbol I}$ is an
identity matrix. By imposing the condition to update whenever
$e^2_k[i] \geq \gamma_k^{2}[i]$ we arrive at the set of all
${\boldsymbol w}_{k}$ that satisfy
\begin{equation}
\sqrt{1-\gamma_k[i]} \leq |{\boldsymbol w}_{k}^{H}[i+1]{\boldsymbol
r}[i]|\leq \sqrt{1+\gamma_k[i]}
\end{equation}
It can be verified that the set above is non-convex and comprises
two parallel hyper-strips in the parameter space.} From the above
conditions we consider two cases: i) $|{\boldsymbol
w}_{k}^{H}[i+1]{\boldsymbol r}[i]| \leq \sqrt{1+\gamma_k[i]}$ and
ii) $|{\boldsymbol w}_{k}^{H}[i+1]{\boldsymbol r}[i]| \geq
\sqrt{1-\gamma_k[i]}$. By substituting the recursion obtained in
(\ref{gradrec}) into $|{\boldsymbol w}_{k}^{H}[i+1]{\boldsymbol
r}[i]|$, we have
\begin{equation}
|{\boldsymbol w}_{k}^{H}[i+1]{\boldsymbol r}[i]|  = |z_{k}[i] -
\mu_{k}[i]e_{k}[i]z_{k}[i]{\boldsymbol
r}^{H}[i]\boldsymbol{\Pi}_{k}[i]{\boldsymbol r}[i]|. \label{eqss}
\end{equation}
Using the above expression we have for case i):
\begin{equation}
|z_{k}[i] - \mu_{k}[i]e_{k}[i]z_{k}[i]{\boldsymbol
r}^{H}[i]\boldsymbol{\Pi}_{k}[i]{\boldsymbol r}[i]| =
\sqrt{1+\gamma_k[i]}
\end{equation}
which leads to
\begin{equation}
\mu_{k}[i]= \Bigg( \frac{1 - \sqrt{1+\gamma_k[i]}}{|z_{k}[i]|}
\Bigg) \frac{1}{e_{k}[i]{\boldsymbol
r}^{H}[i]\boldsymbol{\Pi}_{k}[i]{\boldsymbol r}[i] }
\end{equation}
Using (\ref{eqss}) we have for case ii):
\begin{equation}
|z_{k}[i] - \mu_{k}[i]e_{k}[i]z_{k}[i]{\boldsymbol
r}^{H}[i]\boldsymbol{\Pi}_{k}[i]{\boldsymbol r}[i]| =
\sqrt{1-\gamma_k[i]}
\end{equation}
which results in the following
\begin{equation}
\mu_{k}[i]= \Bigg( \frac{1 - \sqrt{1-\gamma_k[i]}}{|z_{k}[i]|}
\Bigg) \frac{1}{e_{k}[i]{\boldsymbol
r}^{H}[i]\boldsymbol{\Pi}_{k}[i]{\boldsymbol r}[i] }
\end{equation}
The resulting SM-CCM-SG algorithm is described by
\begin{equation}
{\boldsymbol w}_{k}[i+1] = \boldsymbol{\Pi}_{k}[i]({\boldsymbol
w}_{k}[i] - \mu_{k}[i]e_{k}[i]{z}_{k}^{*}[i]{\boldsymbol r}[i]) +
\nu {\boldsymbol p}_k[i]({\boldsymbol p}_k^H[i] {\boldsymbol
p}_k[i])^{-1}, \label{sgalg}
\end{equation}
where
\begin{equation}
\mu_{k}[i] = \left\{ \begin{array}{ll} \Bigg( \frac{1 -
\sqrt{1+\gamma_k[i]}}{|z_{k}[i]|} \Bigg)
\frac{1}{e_{k}[i]{\boldsymbol r}^{H}[i]\boldsymbol{\Pi}_{k}[i]{\boldsymbol r}[i] } & \textrm{if $|z_{k}[i]| \geq\sqrt{(1+\gamma_k[i]}$}\\
\Bigg( \frac{1 - \sqrt{1-\gamma_k[i]}}{|z_{k}[i]|} \Bigg)
\frac{1}{e_{k}[i]{\boldsymbol r}^{H}[i]\boldsymbol{\Pi}_{k}[i]{\boldsymbol r}[i] } & \textrm{if $|z_{k}[i]|\leq\sqrt{(1-\gamma_k[i]}$}\\
0 & \textrm{otherwise}\\
\end{array}\right. \label{step1}
\end{equation}
The SM-CCM-SG algorithm described in (\ref{sgalg})-(\ref{step1})
requires  {$M  + {\rm UR}( 5M - 1)$ additions and $M +1 +{\rm UR}(
5M +4)$} multiplications per received symbol, where ${\rm UR}$ is
the update rate.

\subsection{Set-Membership CCM RLS-Type Algorithm}

In this part, we derive the set-membership CCM recursive
least-squares-type (SM-CCM-RLS) algorithm. The idea is to devise a
least-squares method to calculate a parameter vector ${\boldsymbol
w}_{k}$ used at the receiver for user $k$ that minimizes the
exponentially-weighted CM cost function subject to constraints that
require that the squared error of the filter exceed a specified
time-varying error bound $\gamma_k^2[i]$. { Mathematically, the
proposed SM-CCM-RLS algorithm solves the optimization problem
\begin{equation}
\begin{split}
{\rm minimize}~~~  J_{CM}^{LS}({\boldsymbol w}_{k}[i])  & =
\sum_{l=1}^{i} \lambda_k^{i-l}[i](|{\boldsymbol
w}_{k}^{H}[i]{\boldsymbol r}[l]|^{2}
-1)^{2}   \\
{\rm subject}~~{\rm to} ~~~ {\boldsymbol w}_{k}^H[i] {\boldsymbol
p}_k[i] & = \nu \\ {\rm whenever} ~~ e^2_k[i] & \geq
\gamma_k^{2}[i],
\end{split}
\end{equation}
where $\lambda_k[i]$ is a time-varying forgetting factor. Similarly
to the case of the SM-CCM-SG algorithm, this problem can be solved
with the method of Lagrange multipliers \cite{diniz} along with the
use of the condition $e^2_k[i]  \geq \gamma_k^{2}[i]$ to save
computations
\begin{equation}
\begin{split}
{\mathcal L}({\boldsymbol w}_k[i], \epsilon_k) & = \sum_{l=1}^{i}
\lambda_k^{i-l}[i](|{\boldsymbol
w}_{k}^{H}[i]{\boldsymbol r}[l]|^{2} -1)^{2} \\
& \quad + [ \epsilon_k^*({\boldsymbol p}_k^H[i]{\boldsymbol w}_k[i]
- \nu)], \label{lag2}
\end{split}
\end{equation}
where $\epsilon_k$ is a Lagrange multiplier. Computing the gradient
terms of the Lagrangian in (\ref{lag2}) and equating them to zero,
we obtain
\begin{equation}
\begin{split}
\nabla_{{\boldsymbol w}_k^*[i]}{\mathcal L}({\boldsymbol w}_k[i],
\epsilon_k) & = \sum_{l=1}^{i} \lambda_k^{i-l}[i](|z_k[l]|^{2}
-1)({\boldsymbol r}[i]{\boldsymbol r}^H[i]{\boldsymbol w}_k[i]  + {\boldsymbol p}_k[i] \epsilon_k   = 0 \\
\nabla_{\epsilon_k^*}{\mathcal L}({\boldsymbol w}_k[i], \epsilon_k)
& = {\boldsymbol w}_k^H[i]{\boldsymbol p}_k[i] - \nu = 0.
\label{systeqls}
\end{split}
\end{equation}
Solving for the above equations, we have
\begin{equation}
\begin{split}
{\boldsymbol w}_{k}[i+1] & = \hat{\boldsymbol
R}^{-1}_{k}[i]\Big[\hat{\boldsymbol d}_{k}[i]-\Big({\boldsymbol
p}_{k}^{H}[i]\hat{\boldsymbol R}^{-1}_{k}[i]{\boldsymbol
p}_{k}[i]\Big)^{-1}
\\ &\quad \cdot \Big({\boldsymbol
p}_{k}^{H}[i]\hat{\boldsymbol R}^{-1}_{k}[i] \hat{\boldsymbol
d}_{k}[i] -\nu\Big){\boldsymbol p}_{k}[i]\Big], \label{ccmlsexp}
\end{split}
\end{equation}
where $z_{k}[i]={\boldsymbol w}_{k}^{H}[i]{\boldsymbol r}[i]$,
$\hat{\boldsymbol
R}_{k}[i]=\sum_{l=1}^{i}\lambda_k^{i-l}[i]|z_{k}[l]|^{2}{\boldsymbol
r}[l]{\boldsymbol r}^{H}[l]$, $\hat{\boldsymbol
d}_{k}[i]=\sum_{l=1}^{i}\lambda_k^{i-l}[i]z_{k}^{*}[l]{\boldsymbol
r}[l]$. Firstly, we need to compute $\hat{\boldsymbol d}_{k}[i]$ and
this is performed by the following recursion
\begin{equation}
\hat{\boldsymbol d}_{k}[i] =  \hat{\boldsymbol d}_{k}[i-1] +
\lambda_k[i] z_k^*[i]{\boldsymbol r}[i].
\end{equation}
At this point, we need to compute $\hat{\boldsymbol R}_{k}^{-1}[i]$
efficiently and this is done by applying the matrix inversion lemma
\cite{diniz}, which yields
\begin{equation}
\hat{\boldsymbol R}_{k}^{-1}[i] = \hat{\boldsymbol R}_{k}^{-1}[i-1]
- \frac{\lambda_k[i] |z_k[i]|^2 \hat{\boldsymbol R}_{k}^{-1}[i]
{\boldsymbol r}[i]{\boldsymbol r}^H[i]\hat{\boldsymbol
R}_{k}^{-1}[i-1]}{1+ \lambda_k[i] |z_k[i]|^2  {\boldsymbol
r}^H[i]\hat{\boldsymbol R}_{k}^{-1}[i-1]{\boldsymbol r}[i]}
\label{pinv}
\end{equation}
The last step of the SM-CCM-RLS algorithm is the use of the
condition $e^2_k[i]  \geq \gamma_k^{2}[i]$ to save computations and
to adjust the optimal $\lambda_k[i]$.} In order to adjust
$\lambda_k[i]$, the authors in \cite{nagaraj} have advocated a
strategy that yields bounding ellipsoids that lead to a simple
innovation check with linear complexity, which considers the cost
function
\begin{equation}
{\mathcal{C}}_{\lambda_k[i] } = \lambda_k[i] \Bigg[
\frac{e_k[i]}{\gamma^2_k[i]} \Bigg( \frac{1}{1+
 \lambda_k[i]{\boldsymbol r}^H[i]\hat{\boldsymbol R}_{k}^{-1}[i-1]{\boldsymbol r}[i]} \Bigg) -1 \Bigg]. \label{costl}
\end{equation}
The maximization of the cost function in (\ref{costl}) leads to the
innovation check of the proposed SM-CCM-RLS algorithm:
\begin{equation}
\lambda_k[i] = \left\{ \begin{array}{ll} \frac{1}{{\boldsymbol
r}^H[i]\hat{\boldsymbol R}_{k}^{-1}[i-1]{\boldsymbol r}[i]}
\Bigg( \frac{|e_k[i]|}{\gamma_k[i]} - 1 \Bigg) & \textrm{if $|e^{*}_k[i]|>\gamma_k[i],$}\\
0 & \textrm{otherwise.}\\
\end{array}\right. \label{lamb}
\end{equation}
The SM-CCM-RLS algorithm described in (\ref{ccmlsexp})-(\ref{pinv})
and (\ref{lamb}) requires  {$M  + {\rm UR}( 4M^2 + 3M)$ additions
and $M +2 +{\rm UR}(4M^2+6M +3)$} multiplications per received
symbol. It is worth mentioning that the computational savings can be
quite substantial if the algorithm operates with a low ${\rm UR}$.

\section{Blind Parameter Estimation and Time-varying Bounds}

This section presents a blind framework employed to compute
time-varying error bounds $\gamma_k[i]$ based on parameter and
interference dependency. The proposed blind framework is an
extension of the approach in \cite{delamaresmf} that computes
time-varying error bounds, and performs interference estimation and
tracking. In contrast to \cite{delamaresmf} that considers
training-based recursions, a blind procedure for estimating MAI and
ISI power levels is presented with a set-membership blind channel
estimator and a blind amplitude estimator, which are employed in the
adaptive error bound for the SM adaptive algorithms.

\subsection{Parameter Dependent Bound}

Here, we describe a parameter dependent bound (PDB), that is similar
to the one proposed in \cite{guo} and considers the evolution of the
parameter vector ${\boldsymbol w}_{k}[i]$ for the desired user (user
$k$). The PDB recursion computes a bound for SM adaptive algorithms
and is described by
\begin{equation} \gamma_k[i+1] = (1-\beta)  \gamma_k[i] +
\beta \sqrt{\alpha||{\boldsymbol w}_{k}[i]||^{2}
{\hat{\sigma}}^{2}[i]}, \label{pdb}
\end{equation}
where $\beta$ is a forgetting factor that should be adjusted to
ensure an appropriate time-averaged estimate of the evolution of {
the power of the parameter vector ${\boldsymbol w}_k[i]$. The
quantity $\alpha||{\boldsymbol w}_{k}[i]||^{2}
{\hat{\sigma}}^{2}[i]$ is the variance of the inner product of
${\boldsymbol w}_{k}[i]$ with ${\boldsymbol n}[i]$ which provides
information on the evolution of ${\boldsymbol w}_{k}[i]$, where
$\alpha$ is a tuning parameter and ${\hat{\sigma}}^{2}[i]$ is an
estimate of the noise power. This kind of recursion helps avoiding
too high or low values of the squared norm of ${\boldsymbol w}_k[i]$
and provides a smoother evolution of its trajectory for use in the
time-varying bound. The noise power at the receiver should be
estimated via a time average recursion. In this work, we will assume
that it is known at the receiver.

\subsection{Parameter and Interference Dependent Bound}

In this part, we develop a blind interference estimation and
tracking procedure to be combined with a parameter dependent bound
and incorporated into a time-varying error bound for SM recursions.
The MAI and ISI power estimation scheme, outlined in Fig.
\ref{scheme}, employs both the RAKE receiver and the linear receiver
described in (3) for subtracting the desired user signal from
${\boldsymbol r}[i]$ and estimating MAI and ISI power levels. With
the aid of adaptive algorithms, we design the linear receiver,
estimate the channel modeled as an FIR filter for the RAKE receiver
and obtain the detected symbol $\hat{ b}_{k}[i]$, which is combined
with an amplitude estimate $\hat{A}_{k}[i]$ for subtracting the
desired signal from the output $x_{k}[i]$ of the RAKE. Then, the
difference $d_{k}[i]$ between the desired signal and $x_{k}[i]$ is
used to estimate the MAI and ISI power.

\begin{figure}[!htb]
\begin{center}
\def\epsfsize#1#2{1\columnwidth}
\epsfbox{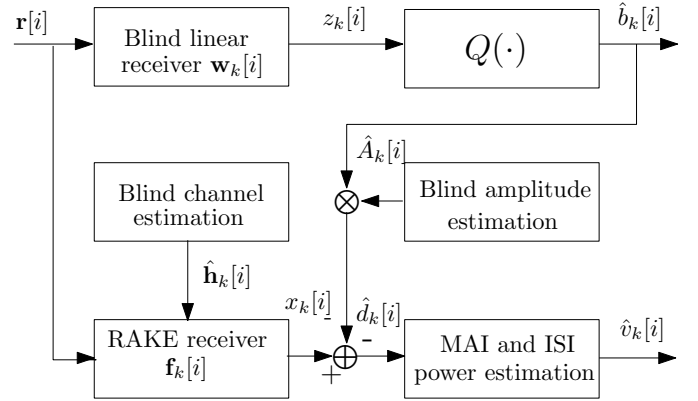} \caption{Block diagram of the proposed blind
scheme.} \label{scheme}
\end{center}
\end{figure}

\subsection{Blind Interference Estimation and Tracking}

{  Let us consider the RAKE receiver with perfect channel knowledge,
whose parameter vector ${\boldsymbol f}_{k}[i]= {\boldsymbol
C}_{k}{\boldsymbol h}_k[i]$ for user $k$ (desired one) corresponds
to the effective signature sequence at the receiver, i.e.
$\tilde{\boldsymbol c}_{k}[i] = {\boldsymbol C}_{k} {\boldsymbol
h}_k[i]$}. The output of the RAKE receiver is given by
\begin{equation}
\begin{split}
x_{k}[i]  = {\boldsymbol f}_{k}^{H}[i]{\boldsymbol r}[i] & =
\underbrace{A_{k}[i] b_{k}[i] {\boldsymbol
f}_{k}^{H}[i]\tilde{\boldsymbol c}_{k}[i]}_{\rm desired ~signal} +
\underbrace{\sum_{\substack{j=2
\\ j \neq k}}^{K}
A_{j}[i]b_{j}[i] {\boldsymbol f}_{j}^{H}[i]\tilde{\boldsymbol c}_{j}[i]}_{\rm MAI} \\
& \quad + \underbrace{{\boldsymbol
f}_{k}^{H}[i]\boldsymbol{\eta_k}[i]}_{\rm ISI} +
\underbrace{{\boldsymbol f}_{k}^{H}[i]{\boldsymbol n}[i]}_{\rm
noise}, \label{xsig}
\end{split}
\end{equation}
where ${\boldsymbol f}_{k}^{H}[i]\tilde{\boldsymbol
c}_{k}[i]=\rho_k[i]$ and ${\boldsymbol
f}_{k}^{H}[i]\tilde{\boldsymbol c}_{j}[i]= \rho_{1,j}[i]$ for $j
\neq 1$. The symbol $\rho_k$ represents the cross-correlation (or
inner product) between the effective signature and the RAKE with
perfect channel estimates. The symbol $\rho_{1,j}[i]$ represents the
cross-correlation between the RAKE receiver and the effective
signature of user $j$. The second-order statistics of the output of
the RAKE in (\ref{xsig}) are described by
\begin{equation}
\begin{split}
E[|x_{k}[i]|^{2}] & = A_{k}^2[i] \rho_k^2[i]
\underbrace{E[|b_{k}[i]|^2]}_{\rightarrow 1} \\ & \quad +
\underbrace{\sum_{\substack{j=1 \\ j \neq k}}^{K}
~~\sum_{\substack{l=1 \\ j \neq k}}^{K}
A_{j}^2[i]E[b_{j}[i]b_{l}^{*}[i]] {\boldsymbol
f}_{j}^{H}\tilde{\boldsymbol c}_{j}\tilde{\boldsymbol
c}_{l}^{H}{\boldsymbol f}_{j}}_{\rightarrow \sum_{j=1,j \neq
k}^{K}{\boldsymbol
f}_{j}^{H}\tilde{\boldsymbol s}_{j}\tilde{\boldsymbol c}_{j}^{H}[i]{\boldsymbol f}_{j}} \\
& \quad + {{\boldsymbol f}_{k}^{H}
E[\boldsymbol{\eta_k}[i]\boldsymbol{\eta_k}^{H}[i]]{\boldsymbol
f}_{k}} + \underbrace{{\boldsymbol f}_{k}^{H}E[{\boldsymbol
n}[i]{\boldsymbol n}^{H}[i]]{\boldsymbol f}_{k}}_{\rightarrow
\sigma^2{\boldsymbol f}_{k}^{H}{\boldsymbol f}_{k}}.
\end{split}
\end{equation}
From the previous development, we can identify the sum of the power
levels of MAI, ISI and noise terms from the second-order statistics.
Our approach is to obtain instantaneous estimates of the MAI, the
ISI and the noise from the output of a RAKE receiver, subtract the
detected symbol in (\ref{symbol}) from this output (using the more
reliable linear multiuser receiver (${\boldsymbol w}_{k}[i]$)) and
to track the interference (MAI + ISI + noise) power as shown in Fig.
1. Let us define the difference between the output of the RAKE
receiver and the detected symbol for user $1$:
\begin{equation}
\begin{split}
d_{k}[i]  = x_{k}[i] - \hat{A}_{k}[i]\hat{b}_{k}[i] & \approx
\underbrace{\sum_{k=2}^{K} A_{k}[i]b_{k}[i] {\boldsymbol
f}_{k}^{H}[i]\tilde{\boldsymbol s}_{k}[i]}_{\rm MAI} +
\underbrace{{\boldsymbol f}_{k}^{H}[i]\boldsymbol{\eta_k}[i]}_{\rm
ISI} \\ & \quad + \underbrace{{\boldsymbol f}_{k}^{H}[i]{\boldsymbol
n}[i]}_{\rm noise}. \label{drec}
\end{split}
\end{equation}
By taking expectations on $|d_{k}[i]|^2$ and taking into account
the assumption that MAI, ISI and noise are uncorrelated we have:
\begin{equation}
\begin{split}
E[|d_{k}[i]|^{2}] & \approx  \sum_{k=2}^{K}{\boldsymbol
f}_{k}^{H}[i]\tilde{\boldsymbol c}_{k}[i]\tilde{\boldsymbol
c}_{k}^{H}[i]{\boldsymbol f}_{k}[i] \\ & \quad + {{\boldsymbol
f}_{k}^{H}[i]
E[\boldsymbol{\eta_k}[i]\boldsymbol{\eta_k}^{H}[i]]}{\boldsymbol
f}_{k}[i] + \sigma^2{\boldsymbol f}_{k}^{H}[i]{\boldsymbol
f}_{k}[i], \label{dsrec}
\end{split}
\end{equation}
where the above equation represents the interference power. Based on
time averages of the instantaneous values of the interference power,
we consider the following algorithm to estimate and track
$E[|d_{k}[i]|^{2}]$
\begin{equation}
\hat{v}[i+1] = (1-\beta) \hat{v}[i] +  \beta |d_{k}[i]|^2,
\label{vrec}
\end{equation}
where $\beta$ is a forgetting factor.  To incorporate parameter
dependency and interference power for computing a more effective
bound {  as an alternative to replace (\ref{pdb})}, we employ the
parameter and interference dependent bound (PIDB)
\begin{equation}
\gamma_k[i+1] = (1-\beta) \gamma_k[i] + \beta \Big(\sqrt{ \tau
~{\hat v}^2[i]} + \sqrt{\alpha||{\boldsymbol w}_{k}||^{2}
{\hat{\sigma}}^{2}[i]}\Big), \label{grec}
\end{equation}
where ${\hat v}[i]$ is the estimated interference power in the
multiuser system and $\tau$ is a weighting parameter that must be
set. The equations in (\ref{vrec}) and (\ref{grec}) are
time-averaged recursions that are aimed at tracking the quantities
$|d_k[i]|^2$ and $(\sqrt{ \tau \hat{v}^2[i]} + \sqrt{\alpha
||{\boldsymbol w}_k ||^2 \hat{\sigma}^2[i]} )$, respectively. The
equations in (\ref{vrec}) and (\ref{grec}) also avoid undesirable
too high or low instantaneous values which may lead to
inappropriate time-varying bound $\gamma_k[i]$.\\

\subsection{Blind Channel and Amplitude Estimation}

Let us now present a set-membership blind channel estimation
(SM-BCE) algorithm to design the RAKE and linear receivers. Consider
the constraint matrix ${\boldsymbol C}_{k}$ that contains one-chip
shifted versions of the signature sequence for user $k$ defined in
(2) and the assumption that the symbols $b_k[i]$ are independent and
identically distributed (i.i.d), and statistically independent from
the symbols of the other users. Consider the covariance matrix
${\boldsymbol R}=[{\boldsymbol r}[i] {\boldsymbol r}^H[i]]$ and the
transmitted signal ${\boldsymbol x}_{k}[i] = {
A}_k[i]{b}_k[i]{\boldsymbol p}_k[i]$, where ${\boldsymbol
p}_k[i]={\boldsymbol C}_k{\boldsymbol h}_k[i]$. Let us now perform
an eigen-decomposition on ${\boldsymbol R}$
\begin{equation}
\begin{split}
{\boldsymbol R} & = \sum_{k=1}^{K} E[{\boldsymbol
x}_k[i]{\boldsymbol x}_k^H[i]] + E[{\boldsymbol \eta}_k[i]
{\boldsymbol \eta}^H_k[i]] + \sigma^2 {\boldsymbol I} \\
& = \big[ {\boldsymbol \phi}_s ~{\boldsymbol \phi}_n \big]
\left[\begin{array}{cc} {{\boldsymbol \Lambda}_s + \sigma^2
{\boldsymbol I}} & {\boldsymbol 0} \\  {\boldsymbol 0} & {\sigma^2
{\boldsymbol I}} \end{array}\right]  \big[ {\boldsymbol \phi}_s
~{\boldsymbol \phi}_n \big]^H,
\end{split}
\end{equation}
where ${\boldsymbol \phi}_s$ and ${\boldsymbol \phi}_n$ are the
signal and noise subspaces, respectively. Since ${\boldsymbol
\phi}_s$ and ${\boldsymbol \phi}_n$ are orthogonal, we have the
condition ${\boldsymbol \phi}_n^H {\boldsymbol x}_{k}[i] =
{\boldsymbol \phi}_n^H {A}_k[i]{b}_k[i]{\boldsymbol p}_k[i] =
{\boldsymbol \phi}_n^H { A}_k[i]{b}_k[i]{\boldsymbol
C}_k{\boldsymbol h}_k[i]={\boldsymbol 0}$. Hence, we have
\begin{equation}
\begin{split}
\Gamma  = {\boldsymbol h}_k^H[i] \underbrace{{\boldsymbol C}_k^H
{b}_k^*[i] { A}_k^*[i] {\boldsymbol \phi}_n {\boldsymbol \phi}_n^H
{A}_k[i] {b}_k[i] {\boldsymbol C}_k}_{\boldsymbol \Upsilon_k}
{\boldsymbol h}_k[i]
\end{split}
\end{equation}
The above relation  allows to blindly estimate the channel
${\boldsymbol h}_k[i]$. To this end, we need to compute the
eigenvector corresponding to the smallest eigenvalue of
${\boldsymbol \Upsilon}_k$. It turns out that we can use the fact
that $\lim_{p \rightarrow \infty} ({\boldsymbol R}/\sigma^2)^{-p}
= {\boldsymbol \phi}_n {\boldsymbol \phi}_n^H$
\cite{delamareccmmimo} and, in practice, it suffices to use $p=1
~{\rm or}~2$. Therefore, to blindly estimate the channel of user
$k$ in the DS-CDMA system we need to solve the optimization
problem
\begin{equation}
\hat{\boldsymbol h}_k[i] = \arg \min_{{\boldsymbol h}_k[i]}
{\boldsymbol h}_k^H[i] {\boldsymbol \Upsilon_k} {\boldsymbol
h}_k[i], ~~ {\rm subject}~~{\rm to} ~~ ||{\boldsymbol h}_k[i]||=1,
\label{cest}
\end{equation}
In order to solve (\ref{cest}) efficiently, we rely on the SM
estimation strategy and a variant of the power method
\cite{delamareccm} that uses a simple shift is adopted to yield the
SM-BCE
\begin{equation}
\hat{\boldsymbol h}_k[i] = ({\boldsymbol I} - \tau_k[i]
\hat{\boldsymbol \Upsilon}_k[i] ) \hat{\boldsymbol h}_k[i-1],
\label{cestpow}
\end{equation}
where $\tau_k[i] = 1/tr[\hat{\boldsymbol \Upsilon}_k[i]]$ and
$\hat{\boldsymbol h}_k[i] \leftarrow \hat{\boldsymbol h}_k[i]/
||\hat{\boldsymbol h}_k[i]||$ to normalize the channel. The
quantity $\hat{\boldsymbol \Upsilon}_k[i]$ is estimated by
\begin{equation}
\hat{\boldsymbol \Upsilon}_k[i] =  \hat{\boldsymbol \Upsilon}_k[i-1]
+ \lambda_k[i]{\boldsymbol C}_k^H \hat{ A}_k^*[i] \hat{\boldsymbol
P}^{p}_k[i] \hat{A}_k[i]  {\boldsymbol C}_k,
\end{equation}
where $\lambda_k[i]$ is a variable forgetting factor that is
obtained by (\ref{lamb}) and $\hat{\boldsymbol P}^{p}_k[i]$ is
computed according to (\ref{pinv}). Next, we describe a procedure
to estimate the amplitude.

In general, amplitude estimation is an important task at the
receiver that is useful for interference cancellation and power
control. The proposed blind interference estimation and tracking
algorithm needs some form of amplitude estimation in order to
accurately compute the interference power. To estimate the
amplitudes of the associated user signals, we describe the following
procedure to estimate the absolute value of the output of the RAKE
receiver defined in (\ref{xsig}) as given by
\begin{equation}
q_k[i+1] = (1-\beta) q_k[i] + |x_k[i]|.
\end{equation}
The amplitude can be estimated by removing the square-root of the
interference power from the above estimate according to
\begin{equation}
\hat A_k[i+1] = (1-\beta) \hat A_k[i] + (|q_k[i]|-|v_k[i]|)
\end{equation}
The above procedure is simple and effective to estimate the amplitude
for use in the interference power estimation procedure. Since one
recursion depends on the other, a designer shall start the procedure
with an interference power equal to zero (or equivalently $|v_k[i]|=0$).

\section{Analysis of The Algorithms}

In this section, we study of the properties of the optimization
problems associated with the design of the SM-CCM-based algorithms
and examine the convergence and tracking performances of the
proposed algorithms.  {In order to ensure the convergence of the
SM-CCM-RLS a persistence of excitation condition on the received
data ${\boldsymbol r}[i]$ must hold and the transmitted symbols
$b_k[i]$ and the noise ${\boldsymbol n}[i]$ have to be uncorrelated.
To this end, the variable bound must adapt such that the above
conditions are met, following the same procedure established for OBE
algorithms in \cite{dasgupta}.} Since the SM-CCM-RLS algorithm is
expected to converge without misadjustment to the Wiener filter and
certain aspects of the convergence analysis of the SM-CCM-SG
algorithm are of greater interest, we will focus on the latter. The
CCM-based algorithms are inherently nonlinear and deal with
time-varying environments, which leads to difficulties in the study
of their performance. For this reason, we will resort to an
effective approach termed energy conservation principle \cite{mai}
that lends itself to such analysis.

\subsection{Analysis of the Optimization Problem}

Let us consider the optimization problem that needs to be solved for
the design of the blind receiver and can be solved by the algorithms
proposed in Section III: {
\begin{equation}
\begin{split}
{\rm minimize} ~ J_{CM}({\boldsymbol w}_k[i]) &
 = E[ ({\boldsymbol w}_k^H[i]{\boldsymbol r}[i]|^2-1)^2]\\ & = E[|z_k[i]|^4- 2|z_k[i]|^2 +1] \\
{\rm subject}~{\rm to} ~ {\boldsymbol w}_k^H[i]{\boldsymbol p}_k[i]
&= \nu \\
{\rm whenever}~~ (|{\boldsymbol w}_k^H[i]{\boldsymbol
p}_k[i]|^2-1)^2 & \geq \gamma_k^2[i] \label{opt_prob}
\end{split}
\end{equation}}
Let us rewrite the received vector as
\begin{equation}
{\boldsymbol r}[i] =\sum_{k=1}^{K} \underbrace{A_k b_k[i]
{\boldsymbol p}_k[i] }_{{\boldsymbol x}_k[i]} + {\boldsymbol
\eta}_k[i] + {\boldsymbol n}[i],
\end{equation}
where ${\boldsymbol p}_k[i] = {\boldsymbol C}_k {\boldsymbol
h}_k[i]$, ${\boldsymbol R} = {\boldsymbol S} + {\boldsymbol G} +
\sigma^2 {\boldsymbol I}$, ${\boldsymbol
S}=E[\sum_{k=1}^{K}{\boldsymbol x}_k[i]{\boldsymbol x}_k^H[i]]$, and
${\boldsymbol G}=E[\sum_{k=1}^{K}{\boldsymbol \eta}_k[i]{\boldsymbol
\eta}^H_k[i]]$, $b_k[i]$ are independent and identically distributed
random variables, and are statistically independent from
${\boldsymbol n}[i] $.

Consider user $k$ as the desired user and let ${\boldsymbol w}_k[i]$
be the receive filter for this user. Our strategy to analyze the
above optimization problem and its properties is to transform the
variables and rewrite the problem in a convenient form that provides
more insight about the nature of the problem. Let us now define the
signal of the desired and the signal of all the users after applying
the receive filter ${\boldsymbol w}_k[i]$:
\begin{equation}
t_k \triangleq A_k {\boldsymbol p}_k^H {\boldsymbol w}_k[i], ~~ {\rm
and}~~ {\boldsymbol t} \triangleq {\boldsymbol A}{\boldsymbol P}^H
{\boldsymbol w}_k = [ t_1 \ldots t_K]^T,
\end{equation}
where ${\boldsymbol P} \triangleq [{\boldsymbol p}_1 \ldots
{\boldsymbol p}_K]$, ${\boldsymbol A} \triangleq {\rm diag} [A_1
\ldots A_K]$ and ${\boldsymbol b} \triangleq [b_1 \ldots b_K]^T$.
The relation between the receive filter, the channel and the
signature sequence can be written as
\begin{equation}
{\boldsymbol w}^H_k[i]{\boldsymbol p}_k[i] = {\boldsymbol
w}^H_k[i]{\boldsymbol C}_k {\boldsymbol h}_k[i] =\nu.
\end{equation}
Therefore, we have for the desired user $k$ the following equivalences
\begin{equation}
t_k[i] = A_k {\boldsymbol p}_k^H[i]{\boldsymbol w}_k[i]= A_k
{\boldsymbol h}_k^H[i] \underbrace{{\boldsymbol C}_k^H{\boldsymbol
w}_k[i]}_{{\boldsymbol h}_k[i]} = A_k {\boldsymbol
h}_k^H[i]{\boldsymbol h}_k[i].
\end{equation}
By considering the noise and the ISI negligible we can write the
cost function in (\ref{opt_prob}) as
\begin{equation}
\begin{split}
J_{CM}({\boldsymbol t}) & = E[|z_k[i]|^4- 2|z_k[i]|^2 +1] 
\\ & = E[ ({\boldsymbol t}^H{\boldsymbol b}{\boldsymbol b}^H{\boldsymbol t})^2]
- 2 E[({\boldsymbol t}^H{\boldsymbol b}{\boldsymbol b}^H{\boldsymbol t}] +1 
\\ &  = 8 (\sum_{j=1}^K t_j t_j^*)^2 - 4 \sum_{j=1}^K (t_j t_j^*)^2 - 4 \sum_{j=1}^K t_j t_j^* +1 \\
& = 8 (D + \sum_{j=1}^K t_j t_j^*)^2 - 4D^2 - 4 \sum_{j=2}^K (t_j
t_j^*)^2 \\ & \quad - 4D - 4 \sum_{j=1,j \neq k}^K (t_j t_j^*) +1 ,
\label{costt}
\end{split}
\end{equation}
where ${\boldsymbol t}=f({\boldsymbol w}_k[i])$ is a linear function
of the receive filter ${\boldsymbol w}_k[i]$, the terms multiplying
the summations in the third line of (\ref{costt}) are obtained by
evaluating the expected values in the second line of (\ref{costt})
\cite{kwak}, and $D=t_j t_j^* = \nu^2 |A_k|^2 |\hat{\boldsymbol
h}_k^H[i] {\boldsymbol h}_k[i]|$. { The strategy we employ to
enforce the convexity of the optimization problem relies on the
adjustment of the parameter $\nu$ and a transformation of variables
on the cost function in (\ref{costt}), which will be detailed in
what follows. }

Let us now transform the above cost function taking into account the
constraint ${\boldsymbol w}_k^H[i]{\boldsymbol p}_k[i] = \nu$ and
the fact that we are interested in demodulating user $k$ and
rejecting the remaining users. Therefore, we introduce another
parameter vector  $\bar{\boldsymbol t} = [t_1 \ldots t_{k-1} ~
t_{k+1} \ldots t_K]^T = \bar{\boldsymbol B} {\boldsymbol w}_k[i]$
that excludes the user $k$ and is responsible for the interference
suppression task of the remaining users (all but user $k$), where
$\bar{\boldsymbol B} = \bar{\boldsymbol A}\bar{\boldsymbol P}^H$,
$\bar{\boldsymbol P}[i] = [ {\boldsymbol p}_1[i] \ldots
~{\boldsymbol p}_{k-1}[i] ~ {\boldsymbol p}_{k+1}[i] \ldots
{\boldsymbol p}_K[i]]$ and ${\boldsymbol A} = {\rm diag} [A_1 \ldots
A_{k-1} ~ A_{k+1} \ldots A_K]$. The transformed cost function is
given by
\begin{equation}
\begin{split}
J_{CM}(\bar{\boldsymbol t}) & = 8 (D + \bar{\boldsymbol
t}^H\bar{\boldsymbol t})^2 - 4( D^2 +  \bar{\boldsymbol
t}^H\bar{\boldsymbol t})^2 \\ & \quad - 4(D +\bar{\boldsymbol
t}^H\bar{\boldsymbol t}) +1.
\end{split}
\end{equation}
At this point we need to take into account the constraint
$J_{CM}({\boldsymbol w}_k[i]) \leq \gamma_k^2[i]$. Since
$\bar{\boldsymbol t}=\bar{\boldsymbol B} {\boldsymbol w}_k[i]$ is a
linear  mapping we have an equivalent constraint
$J_{CM}(\bar{\boldsymbol t}) \leq \bar{\gamma}_k^2[i]$, where
$\bar{\gamma}_k^2[i]$ is the bound modified by the linear mapping.
It can be verified that the constraint set generated by
$J_{CM}(\bar{\boldsymbol t}) \leq \bar{\gamma}_k^2[i]$ is not convex
and leads to two disjoint parallel hyperstrips in the parameter
space.

Given that the constraint is not convex, the optimization problem is
clearly non convex. However, in this context non-convexity only
poses a problem if local minima are present and prevent an algorithm
from reaching the global minimum. It turns out that a designer can
adjust the parameter $\nu$ in order to enforce the convexity for
each hyperstrip. Computing the Hessian \cite{bert} ${\boldsymbol H}
= \frac{\partial}{\partial \bar{\boldsymbol t}^H} \frac{\partial
J_{CM}({\boldsymbol t})}{\partial \bar{\boldsymbol t}}$ we obtain
\begin{equation}
\begin{split}
{\boldsymbol H} & = 16 \big[ (D-1/4) {\boldsymbol I} + \\ & \quad
\underbrace{ \bar{\boldsymbol t}^H  \bar{\boldsymbol t} {\boldsymbol
I} + \bar{\boldsymbol t} \bar{\boldsymbol t}^H - {\rm diag} \big(
|t_1[i]|^2 \ldots |t_{k-1}|^2 ~ |t_{k+1}|^2 \ldots |t_K[i]|^2
\big)}_{\rm positive~ or~ positive ~semi-definite~ terms} \big].
\end{split}
\end{equation}
The condition $D=\nu^2 |A_k|^2 |\hat{\boldsymbol h}_k^H[i]
\hat{\boldsymbol h}_k[i]|^2 \geq 1/4$ ensures that there is no local
minimum in each of the hyperstrips resulting from the optimization
problem and convexity can be enforced in each of the hyperstrips.
Since ${\boldsymbol t}=\bar{\boldsymbol B} {\boldsymbol w}_k[i]$ is
a linear function of ${\boldsymbol w}_k[i]$ then
$J_{CM}(\bar{\boldsymbol t})$ preserves the maxima and minima
properties of $J_{CM}({\boldsymbol w}_k[i])$. For sufficiently high
signal-to-noise ratio (SNR) values, the extrema of the cost function
can be considered a small perturbation of the noise-free case
\cite{kwak}.

\subsection{Stability Analysis}

In this part, we discuss the stability analysis of the SM-CCM-SG
algorithm described in subsection III.A. In particular, we consider
the range of step-size values for convergence. Let us now rewrite
the update equation of the algorithm as
\begin{equation}
\begin{split}
{\boldsymbol w}_{k}[i+1] & = \boldsymbol{\Pi}_{k}[i]({\boldsymbol
w}_{k}[i] - \mu_{k}[i]e_{k}[i]{z}_{k}^{*}[i]{\boldsymbol r}[i]) +
\nu {\boldsymbol p}_k[i]({\boldsymbol p}_k^H[i] {\boldsymbol
p}_k[i])^{-1}\\  & = {\boldsymbol w}_{k}[i] -
\mu_{k}[i]e_{k}[i]{\boldsymbol r}^H[i]{\boldsymbol w}_{k}[i] \big(
{\boldsymbol I} - \frac{{\boldsymbol p}_k[i]{\boldsymbol
p}_k^H[i]}{{\boldsymbol p}_k^H[i]{\boldsymbol p}_k[i] }
\big){\boldsymbol r}[i] \\ & = {\boldsymbol w}_{k}[i] -
\mu_{k}[i]e_{k}[i]{\boldsymbol v}[i]{\boldsymbol r}^H[i]{\boldsymbol
w}_{k}[i] \\ & = ({\boldsymbol I} - \mu_{k}[i]e_{k}[i]{\boldsymbol
v}[i]{\boldsymbol r}^H[i]){\boldsymbol w}_{k}[i], \label{algen}
\end{split}
\end{equation}
where ${\boldsymbol v}_k[i] = ({\boldsymbol I} - \frac{{\boldsymbol
p}_k[i] {\boldsymbol p}_k^H[i]}{{\boldsymbol p}_k^H[i]{\boldsymbol
p}_k[i]}{\boldsymbol r}[i])$ and the expression in the second line
of (\ref{algen}) is obtained by substituting ${\boldsymbol \Pi}_k[i]
= {\boldsymbol I} - {\boldsymbol p}_k[i]({\boldsymbol p}_k^H[i]
{\boldsymbol p}_k[i])^{-1} {\boldsymbol p}_k^H[i]$ into the first
line and further manipulating the terms. Let us now define the error
vector
\begin{equation}
\begin{split}
{\boldsymbol \epsilon}_{{\boldsymbol w}_k}[i+1] & = {\boldsymbol w}_{k,{\rm opt}} - {\boldsymbol w}_k[i+1] \\
& = ({\boldsymbol I} - \mu_k[i] e_k[i] {\boldsymbol v}_k[i] {\boldsymbol r}^H[i]) {\boldsymbol \epsilon}_{{\boldsymbol w}_k}[i] + \mu_k[i] e_k[i] {\boldsymbol v}_k[i] {\boldsymbol r}^H[i] {\boldsymbol w}_{k,{\rm opt}}. \label{errorvec}
\end{split}
\end{equation}
In order to proceed with the analysis, we need to resort to an
assumption.

{\it Assumption 1}: Let us suppose that for the algorithm in
(\ref{algen}) when $i \rightarrow \infty$
\begin{equation}
E[\mu_k[i] e_k^2[i]] = E[\mu_k[i]] E[e_k^2[i]].
\end{equation}
This assumption holds if $\mu_k[i]$ is a constant, and we claim that
it is approximately true if $\mu_k[i]$ varies slowly around its mean
value. By writing
\begin{equation}
E[\mu_k[i] e_k^2[i] ] =E[\mu_k[i] ] E[e_k^2[i]] + E[( \mu_k[i] -
E[\mu_k[i]] ) e_k^2[i] ],
\end{equation}
we can notice that the second term on the right-hand side will be
small compared with the first one provided that $\mu_k[i]$ varies
slowly around its mean value.

By taking expectations on both sides of (\ref{errorvec}) and using
the previous assumption, we have
\begin{equation}
E[{\boldsymbol \epsilon}_{{\boldsymbol w}_k}[i+1]] = ({\boldsymbol
I}- E[\mu_k[i]] {\boldsymbol R}_{\boldsymbol rv}) E[ {\boldsymbol
\epsilon}_{{\boldsymbol w}_k}[i]], \label{stab}
\end{equation}
where ${\boldsymbol R}_{\boldsymbol rv}= E[e_k[i] {\boldsymbol
v}_k[i] {\boldsymbol r}^H[i] ]$ and ${\boldsymbol R}_{\boldsymbol
vr} {\boldsymbol w}_{\rm opt} = {\boldsymbol 0}$.

From the above, it can be concluded that ${\boldsymbol w}_k[i]$
converges to ${\boldsymbol w}_{k,{\rm opt}}$ and (\ref{stab}) is
stable if and only if $\prod_{i=0}^{\infty} ( {\boldsymbol I} -
E[\mu_k[i]] {\boldsymbol R}_{\boldsymbol vr}) \rightarrow 0 $, which
is a necessary and sufficient condition for $\lim_{i \rightarrow
\infty} E[{\boldsymbol \epsilon}_{{\boldsymbol w}_k} ] =
{\boldsymbol 0}$ and $E[ {\boldsymbol w}_k[i]] \rightarrow
{\boldsymbol w}_{k,{\rm opt}}$. For stability, a sufficient
condition for (\ref{stab}) to hold implies that
\begin{equation}
0\leq E[ \mu(\infty)] < \min_k \frac{2}{|\lambda_k^{vr}| },
\end{equation}
where $\lambda_k^{vr}$ is the $k$th eigenvalue of ${\boldsymbol
R}_{\boldsymbol vr}$ that is not necessarily real since
${\boldsymbol R}_{\boldsymbol rv}$ is not symmetric.

\subsection{Steady-State Analysis}

In this part of the analysis we are interested in devising a formula
to predict the excess MSE, which depends on the MAI, the ISI, the
noise at the receiver and the parameters of the SM-CCM-SG algorithm.
The excess MSE is related to the error in the filter coefficients
${\boldsymbol \epsilon}_{{\boldsymbol w}_k}[i]$ via the {\it a
priori} estimation error, which is defined as
\begin{equation}
e_a[i] \triangleq {\boldsymbol \epsilon}_{{\boldsymbol w}_k}^H[i]{\boldsymbol r}[i],
\end{equation}
where ${\boldsymbol \epsilon}_{{\boldsymbol w}_k}[i] = {\boldsymbol w}_{\rm opt} - {\boldsymbol w}_k[i]$, and ${\boldsymbol w}_{\rm opt}$ is the optimum linear MMSE receiver. Consider the MSE at time $i$:
\begin{equation}
\begin{split}
{\rm MSE}[i] & = E[ |b_k[i] - {\boldsymbol w}_k^H[i]{\boldsymbol r}[i]|^2] \\
& = \epsilon_{\rm min} + E[|e_a[i]|^2] + {\boldsymbol p}_k^H[i]
E[{\boldsymbol \epsilon}_{{\boldsymbol w}_k}[i]] + E[{\boldsymbol
\epsilon}_{{\boldsymbol w}_k}^H[i]]{\boldsymbol p}_k[i] \\ & \quad -
E[ {\boldsymbol w}_{\rm opt}^H {\boldsymbol r}[i] {\boldsymbol
r}^H[i] {\boldsymbol \epsilon}_{{\boldsymbol w}_k}[i]] -
E[{\boldsymbol \epsilon}_{{\boldsymbol w}_k}^H[i] {\boldsymbol
r}[i]{\boldsymbol r}^H[i] {\boldsymbol w}_{k,{\rm opt}}],
\end{split}
\end{equation}
where {  $e_a[i] \triangleq {\boldsymbol \epsilon}_{{\boldsymbol
w}_k}^H[i] {\boldsymbol r}[i] = ({\boldsymbol w}_{\rm opt} -
{\boldsymbol w}_k[i])^H{\boldsymbol r}[i]$} and $\epsilon_{\rm min}
= E[|b_k[i] - {\boldsymbol w}_{\rm opt}^H{\boldsymbol r}[i]|^2]$.
When $i \rightarrow \infty$, we have ${\boldsymbol w}_k[i]
\rightarrow {\boldsymbol w}_{k,{\rm opt}}$ and $E[{\boldsymbol
\epsilon}_{{\boldsymbol w}_k}[i]] \rightarrow 0$ and the
steady-state MSE
\begin{equation}
\lim_{i \rightarrow \infty} {\rm MSE}[i] =  \epsilon_{\rm min} +
\lim_{i \rightarrow \infty} E[|e_a[i]|^2]
\end{equation}
The steady-state excess MSE is then defined \cite{diniz} as
\begin{equation}
\xi \triangleq   \lim_{i \rightarrow \infty} E[|e_a[i]|^2]
\end{equation}
Using the energy conservation principle \cite{mai}, the proposed
SM-CCM-SG algorithm can be written in the form
\begin{equation}
{\boldsymbol w}_k[i+1] = {\boldsymbol w}_k[i] + \mu_k[i] \underbrace{( -e_k[i] z^*_k[i])}_{F_{e_k}[i]} \underbrace{ \bigg({\boldsymbol I} - \frac{{\boldsymbol p}_k[i] {\boldsymbol p}_k^H[i]}{{\boldsymbol p}_k^H[i] {\boldsymbol p}_k[i] }\bigg) {\boldsymbol r}[i]}_{{\boldsymbol u}_k[i]},
\end{equation}
where $F_{e_k}[i]$ is a generic scalar function determined by the
adaptive algorithm. Subtracting the above recursion for
${\boldsymbol w}_k[i+1]$ from ${\boldsymbol w}_{k,{\rm opt}}$, we
obtain
\begin{equation}
{\boldsymbol \epsilon}_{{\boldsymbol w}_k}[i+1] = {\boldsymbol \epsilon}_{{\boldsymbol w}_k}[i] - \mu_k[i] {\boldsymbol u}_k[i] F_{e_k}[i]
\end{equation}
Using the {\it a priori} estimation error $e_a[i] \triangleq
{\boldsymbol \epsilon}_{{\boldsymbol w}_k}^H[i] {\boldsymbol r}[i]$.
Rewriting the previous equation, we obtain
\begin{equation}
\begin{split}
e_a[i]  & = e_p[i] + \mu_k[i] {\boldsymbol u}_k^H[i] {\boldsymbol r}[i]  F_{e_k}^*[i] \\
& = e_p[i] + \mu_k[i]  {\boldsymbol r}^H[i] \bigg({\boldsymbol I} -
\frac{{\boldsymbol p}_k[i] {\boldsymbol p}_k^H[i]}{{\boldsymbol
p}_k^H[i] {\boldsymbol p}_k[i] }\bigg) {\boldsymbol r}[i]
F_{e_k}^*[i]
\end{split}
\end{equation}
Since ${\boldsymbol r}^H[i] \bigg({\boldsymbol I} - \frac{{\boldsymbol p}_k[i] {\boldsymbol p}_k^H[i]}{{\boldsymbol p}_k^H[i] {\boldsymbol p}_k[i] }\bigg) {\boldsymbol r}[i] = {\boldsymbol u}_k^H[i]{\boldsymbol u}_k[i]$, we have
\begin{equation}
e_p[i] = e_a[i] - \mu_k[i]{\boldsymbol u}_k^H[i] {\boldsymbol
u}_k[i] F_{e_k}^*[i]. \label{errorp}
\end{equation}
Using the fact that $F_{e_k}^*[i]= \frac{e_a[i] -e
_p[i]}{\mu_k[i]{\boldsymbol u}_k^H[i] {\boldsymbol u}_k[i]}$, we can
rewrite ${\boldsymbol \epsilon}_{{\boldsymbol w}_k}[i]$ as
\begin{equation}
{\boldsymbol \epsilon}_{{\boldsymbol w}_k}[i+1]={\boldsymbol
\epsilon}_{{\boldsymbol w}_k}[i] - \frac{{\boldsymbol
u}_k[i]}{{\boldsymbol u}_k^H[i]{\boldsymbol u}_k[i]} [ e_a^*[i] -
e_p^*[i]]
\end{equation}
Rearranging, we obtain
\begin{equation}
{\boldsymbol \epsilon}_{{\boldsymbol w}_k}[i+1] + \frac{{\boldsymbol
u}_k[i]}{{\boldsymbol u}_k^H[i]{\boldsymbol u}_k[i]}e_p^*[i]
={\boldsymbol \epsilon}_{{\boldsymbol w}_k}[i] + \frac{{\boldsymbol
u}_k[i]}{{\boldsymbol u}_k^H[i]{\boldsymbol u}_k[i]} e_a^*[i]
\end{equation}
By defining $\bar{\mu}[i] = 1/({\boldsymbol u}_k^H[i]{\boldsymbol
u}_k[i])$, squaring the previous equation and taking expectations,
we obtain the relation
\begin{equation}
E[||{\boldsymbol \epsilon}_{{\boldsymbol w}_k}[i+1]||^2]  +
E[\bar{\mu}_k[i]|e_p[i]|^2] = E[||{\boldsymbol
\epsilon}_{{\boldsymbol w}_k}[i+1]||^2] +
E[\bar{\mu}_k[i]|e_a[i]|^2], \label{relw}
\end{equation}
In the steady state, we can write
\begin{equation}
E[||{\boldsymbol \epsilon}_{{\boldsymbol w}_k}[i]||^2]  = E[
||{\boldsymbol \epsilon}_{{\boldsymbol w}_k}[i+1]||^2], ~~{\rm
and}~~ E[\bar{\mu}_k[i] |e_a[i]|^2] = E[\bar{\mu}_k[i] |e_p[i]|^2]
\end{equation}
{  The energy preserving equation can be used to compute the excess
MSE. It can be obtained by cancelling the terms $E[||{\boldsymbol
\epsilon}_{{\boldsymbol w}_k}[i+1]||^2]$ and $E[||{\boldsymbol
\epsilon}_{{\boldsymbol w}_k}[i]||^2]$ in (\ref{relw}) and by
substituting (\ref{errorp}) into (\ref{relw}), which yields}
\begin{equation}
E[\bar{\mu}_k[i] |e_a[i]|^2] = E[\bar{\mu}_k[i] |e_a[i]-
\frac{\mu_k[i]}{\bar{\mu}_k[i]} F_{e_k}^*[i]|^2], \label{rel}
\end{equation}
where $F_{e_k}[i] = - e_k[i] z_k^*[i] = (1 - |z_k[i]|^2) z_k^*[i]$
and $z_k[i] = ({\boldsymbol w}_{k,{\rm opt}} - {\boldsymbol
\epsilon}_{{\boldsymbol w}_k}[i])^H {\boldsymbol r}[i] =
{\boldsymbol w}_{k,{\rm opt}}^H{\boldsymbol r}[i] - e_a[i] = A_k
b_k[i] + {\rm MAI} + {\boldsymbol \eta}_k[i] + v[i] - e_a[i]$.
Substituting $F_{e_k}[i]$ and (\ref{errorp}) into (\ref{rel}) and
manipulating the terms, we obtain
\begin{equation}
\begin{split}
&E[\mu_k[i]] E[e_a[i] z_k[i] (1 - |z_k[i]|^2)] \\ & \underbrace{+
E[\mu_k[i]] E[\mu_k[i]] E[e_a[i] z_k^*[i] (1-|z_k[i]|^2)]}_{C} \\ &
= \underbrace{E[\mu_k^2[i]] E[{\boldsymbol u}_k^H[i]{\boldsymbol
u}_k[i] |z_k[i]|^2(1-|z_k[i]|^2)^2]}_{F}. \label{fund_rel}
\end{split}
\end{equation}
At this point we need to resort to another assumption to continue with our analysis.

{\it Assumption 2:} In the steady state, ${\boldsymbol
u}_k^H[i]{\boldsymbol u}_k[i]$ and $|F_{e_k}[i]|^2$ are
uncorrelated. The quantities $\{ b_k[i], {\rm MAI}, {\boldsymbol
\eta}[i], v[i], e_a[i] \}$ are zero mean random variables, and are
mutually independent. {  We use the fact that $E[b_k^{2m}]=1$ for
any positive integer $m$ and that the residual MAI and ISI are
Gaussian random variables.}

Using the previous assumption and substituting {
$z_k[i]={\boldsymbol w}_k^H[i]{\boldsymbol r}[i]$} into
(\ref{fund_rel}), we obtain
\begin{equation}
\begin{split}
&E[\mu_k^2[i]]E[{\boldsymbol u}_k^H[i]{\boldsymbol u}_k[i]] A
E[|e_a[i]|^2]\\ & + 3E[\mu_k^2[i]] E[{\boldsymbol
u}_k^H[i]{\boldsymbol u}_k[i]] E[{\rm MAI}^2]   E[|e_a[i]|^4] \\ & +
3E[\mu_k^2[i]] E[{\boldsymbol u}_k^H[i]{\boldsymbol u}_k[i]]
E[v^2[i]] E[|e_a[i]|^4] \\ & + E[\mu^2[i]] E[{\boldsymbol
u}_k^H[i]{\boldsymbol u}_k[i]] (3E[\mu^2[i]] +1 ) E[|e_a[i]|^4] \\ &
\underbrace{E[\mu_k^2[i]] E[{\boldsymbol u}_k^H[i]{\boldsymbol
u}_k[i]]B + E[\mu_k^2[i]] E[{\boldsymbol u}_k^H[i]{\boldsymbol
u}_k[i]] E[|e_a[i]|^6]}_{C}  \\ & =2 E[\mu_k[i]](E[{\rm MAI}^2]
E[|e_a[i]|^2] + E[v^2[i]] E[|e_a[i]|^2]\\ & \quad \underbrace{+
E[{\boldsymbol \eta}^2[i]] E[|e_a[i]|^2] + E[|e_a[i]|^4]}_{D},
\label{fund_rel2}
\end{split}
\end{equation}
where $A=3+3\sigma^4 {\rm MAI} + 6\sigma^2_{\rm MAI}\sigma_v^2 +
6\sigma_{\rm MAI} \sigma_n^2 + 3 \sigma_v^4 + 6 \sigma_v^2
\sigma_n^2 + 3 \sigma_n^4$ and $B = \sigma_{\eta}^6 + 3 \sigma_v^2
\sigma_{\eta}^4 + 3 \sigma_{\rm MAI}^2\sigma_{\eta}^4 + \sigma_v^6 +
6 \sigma_v^2 + \sigma_{\eta}^2 \sigma_{\rm MAI}^2 + 3 \sigma_{\rm
MAI}^4\sigma_n^2 + 3 \sigma_{\rm MAI}^4 \sigma_v^2 + \sigma_{\rm
MAI}^6 + \sigma_{\rm MAI}^4 + 2 \sigma^2_v\sigma_n^2 + \sigma^4_v +
2 \sigma_{\rm MAI}^2 + 2 \sigma_{\rm MAI}^2 \sigma_v^2 + \sigma_{\rm
MAI}^4 + 4 \sigma_v^2 + 2 \sigma_{\eta}^2 + 2 \sigma_{\rm MAI}^2
+2$.

Upon convergence($i \rightarrow \infty$), we can assume that $E[{\rm
MAI}^{2m}] (E[{\rm MAI}^2])^m = \sigma_{\rm MAI}^{2m}$,
$E[\eta^{2m}] = (E[\eta^2])^m = \sigma_{\eta}^{2m}$, and $E[v^{2m}]
= (E[v^2])^m = \sigma_v^{2m}$, where $\sigma_{\rm MAI}$,
$\sigma_{\eta}$, and $\sigma_v$ are the variances of the Gaussian
distribution. In this situation, the high power terms
$E[|e_a[i]|^4]$ and $E[|e_a[i]|^6]$ may be neglected as these values
are typically very small compared with the remaining terms. The
excess MSE is obtained as follows
\begin{equation}
\xi = E[|e_a[i]|^2] = \frac{E[\mu_k^2(\infty)] E[{\boldsymbol
u}_k^H[i] {\boldsymbol u}_k[i]] B}{2E[\mu_k(\infty)] (\sigma_{\rm
MAI}^2 + \sigma_v^2 + \sigma_{\eta}^2) - E[\mu_k^2(\infty)]
E[{\boldsymbol u}_k^H[i]{\boldsymbol u}_k[i]]A} \label{emse_ss}
\end{equation}

\subsection{Tracking Analysis}

In this subsection, we assess the proposed SM-CCM-SG algorithm in a
non-stationary environment, in which the algorithm has to track the
minimum point of the error-performance surface. Specifically, we
derive an expression for the excess MSE of a blind adaptive linear
receiver when the channel varies in time. Differently from the works
in \cite{mai,whitehead}, where expressions were derived for the
constant modulus (CM) algorithm and the CCM algorithm, respectively,
we consider a set-membership approach. In the time-varying scenarios
of interest, the optimum receive filter coefficients are assumed to
vary according to the model ${\boldsymbol w}_{\rm opt}[i+1]  =
{\boldsymbol w}_{\rm opt} + {\boldsymbol q}[i]$, where ${\boldsymbol
q}[i]$ denotes a random perturbation. This is consistent with
tracking analyses of adaptive filtering algorithms and requires an
assumption.

{\it Assumption 3:} The sequence ${\boldsymbol q}[i]$ is a
stationary sequence of independent zero-mean vectors and positive
definite autocorrelation matrix ${\boldsymbol Q}=E[{\boldsymbol
q}[i] {\boldsymbol q}^H[i]]$, which is mutually independent of the
sequences $\{ {\boldsymbol u}_k[i] \}$, $\{ v[i] \}$, $\{ {\rm
MAI}[i] \}$, $\{ \eta[i]\}$.

Let us consider the weight-error vector ${\boldsymbol
\epsilon}_{{\boldsymbol w}_k}[i] = {\boldsymbol w}_{k,{\rm opt}} -
{\boldsymbol w}_k[i]$, which satisfies
\begin{equation}
{\boldsymbol \epsilon}_{{\boldsymbol w}_k}[i+1] = {\boldsymbol
\epsilon}_{{\boldsymbol w}_k}[i] -\mu_k[i] {\boldsymbol u}_k[i]
F_{e_k}[i] + {\boldsymbol q}[i]. \label{rel1}
\end{equation}
With $e_a[i] = {\boldsymbol \epsilon}_{{\boldsymbol
w}_k}^H[i]{\boldsymbol r}[i]$ and $e_p[i] = {\boldsymbol
\epsilon}_{{\boldsymbol w}_k}^H[i+1]{\boldsymbol r}[i]$, we have
\begin{equation}
e_a[i] = e_p[i] + \mu_k[i] ||{\boldsymbol u}_k[i]||^2 F_{e_k}^*[i]. \label{rel2}
\end{equation}
Using (\ref{rel1}) and (\ref{rel2}), we obtain
\begin{equation}
{\boldsymbol \epsilon}_{{\boldsymbol w}_k}[i+1]+ \bar{\mu}_k[i]
{\boldsymbol u}_k[i] e_a^*[i] = {\boldsymbol \epsilon}_{{\boldsymbol
w}_k}[i] + {\boldsymbol q}[i] + \bar{\mu}_k[i] {\boldsymbol u}_k[i]
e_p^*[i]. \label{rel3}
\end{equation}
Squaring (\ref{rel3}) and taking the expected value on both sides,
we obtain
\begin{equation}
\begin{split}
E[||{\boldsymbol \epsilon}_{{\boldsymbol w}_k}[i+1]||^2] +
E[\bar{\mu}_k[i] |e_a[i]|^2] & = E[||{\boldsymbol
\epsilon}_{{\boldsymbol w}_k}[i]+{\boldsymbol q}[i]||^2 ] \\ & \quad
+ E[\bar{\mu}_k[i] |e_p[i]|^2],
\end{split}
\end{equation}
where
\begin{equation}
\begin{split}
E[||{\boldsymbol \epsilon}_{{\boldsymbol w}_k}[i]+{\boldsymbol
q}[i]||^2] & = E[||{\boldsymbol \epsilon}_{{\boldsymbol
w}_k}[i]||^2] + E[{\boldsymbol \epsilon}_{{\boldsymbol
w}_k}[i]+{\boldsymbol q}[i]] \\ & \quad + E[{\boldsymbol q}^H[i]
{\boldsymbol \epsilon}_{{\boldsymbol w}_k}[i]]+ E[{\boldsymbol
q}^H[i]{\boldsymbol q}[i]].
\end{split}
\end{equation}
Using {\it Assumption 3}, we have $E[{\boldsymbol
\epsilon}_{{\boldsymbol w}_k}^H[i]{\boldsymbol q}[i]]=E[{\boldsymbol
q}^H[i] {\boldsymbol \epsilon}_{{\boldsymbol w}_k}[i]=0$. When $i
\rightarrow \infty$ $E[||{\boldsymbol \epsilon}_{{\boldsymbol
w}_k}[i+1]||^2] = E[||{\boldsymbol \epsilon}_{{\boldsymbol
w}_k}[i]||^2]$, the energy preserving equation describing the
tracking performance is given by
\begin{equation}
E[\bar{\mu}_k[i] |e_a[i]|^2] = {\rm Tr}({\boldsymbol Q}) + E[ \bar{\mu}_k[i] |e_a[i] - \frac{\mu_k[i]}{\bar{\mu}_k[i]} F_{e_k}^*[i]|^2
\end{equation}
Expanding the equation above, it can be simplified to
\begin{equation}
C = {\rm Tr}({\boldsymbol Q}) + D,
\end{equation}
where $C$ and $D$ are defined in (\ref{fund_rel2}). The excess MSE
is then obtained as
\begin{equation}
\begin{split}
\xi & = E[|e_a[i]|^2] \\ & = \frac{E[ \mu_k^2 (\infty)]
E[||{\boldsymbol u}_k[i]||^2] B + {\rm Tr}( {\boldsymbol Q} )}{ 2 E[
\mu_k (\infty) (\sigma_{\rm MAI}^2 + \sigma_v^2 + \sigma_{\eta}^2 )
- E[\mu^2_k (\infty) ] E[|{\boldsymbol u}_k[i]|^2] A}
\label{emse_tr}
\end{split}
\end{equation}

\subsection{Computation of Moments}

In order to compute the expressions obtained in (\ref{emse_ss}) and
(\ref{emse_tr}), we need to obtain the first and second-order
moments of $\mu_k[\infty]$. To this end, we resort to the expression
for the variable step size in (\ref{step1}) and the methodology
employed in \cite{werner1}, which after some algebraic manipulations
yields
\begin{equation}
E[\mu_k[\infty]] = E[\gamma_k[i]] P_{\rm up} + \frac{(1-P_{\rm
up})}{E[\gamma_k[i]]} \frac{E[A_k[i]]}{E[|{\boldsymbol u}_k[i]|^2]},
\end{equation}
\begin{equation}
E[\mu_k^2[\infty]] = E[\gamma_k[i]] P_{\rm up} + \frac{(1-P_{\rm
up})}{E[\gamma_k[i]]} \frac{E[A_k^2[i]]}{E[|{\boldsymbol
u}_k[i]|^4]},
\end{equation}
where the probability of update $P_{\rm up}$ is given by
\begin{equation}
\begin{split}
 P_{\rm up} & = Pr\big[ | e_k[i] | > E[\gamma_k[i]] \big]  = Pr\big[
 |e_k[i]|^2 > E[|\gamma_k[i]|^2 \big] \\ & = 2 Q
 \bigg(\frac{E[\gamma_k[i]]}{\sigma_e} \bigg),
\end{split}
\end{equation}
where $Pr[\cdot]$ denotes the probability $Q(x)$ is the
complementary Gaussian cumulative distribution function \cite{rappa}
given by
\begin{equation}
Q(x) = \int_{x}^{\infty} \frac{1}{\sqrt{2 \pi}} e^{-t^2/2} dt.
\end{equation}

\section{Simulation Results}

In this section we assess the performance of the proposed and
analyzed adaptive algorithms in terms of mean-square error (MSE) and
bit error rate (BER). In particular, we consider the proposed
SM-CCM-SG and SM-CCM-RLS adaptive algorithms and the existing CMV-SG
and CMV-RLS reported in \cite{xutsa}, the CCM-SG \cite{xu&liu} and
the CCM-RLS \cite{delamareccm} algorithms, the SM-CMV-SG and
SM-CMV-RLS reported in \cite{gollamudi2} with and without the
proposed blind PDB and PIDB time-varying bounds described in Section
IV. The DS-CDMA network employs Gold sequences of length $N=31$, the
users are randomly distributed and communicate over multipath
channels. The channels experienced by different users are
independent and different since we focus on an uplink scenario. The
DS-CDMA system under consideration employs quadrature phase-shift
keying (QPSK) modulation. The channel coefficients are given by
$h_{k,l}[i]=p_{k,l}\alpha_{k,l}[i]$, where $\alpha_{k,l}[i]$
($l=0,1,\ldots,L_{p}-1$, $k=1,2,\ldots, K$) are obtained with
Clarke's model \cite{rappa} and $p_{k,l}$ represent the powers of
each channel path. We show the results in terms of the normalized
Doppler frequency $f_{d}T$ (cycles/symbol), where $f_d$ is the
maximum Doppler shift and $T$ is the symbol interval. We  use
three-path channels with relative powers given by $0$, $-3$ and $-6$
dB,where in each run the spacing between paths is obtained from a
discrete uniform random variable between $1$ and $2$ chips. The
proposed blind channel estimator described in Section IV and that of
\cite{douko} model the channel as a finite impulse response (FIR)
filter and we employ a filter with $6$ taps as an upper bound for
the experiments. The phase ambiguity derived from the blind channel
estimation method in \cite{douko} is addressed in our simulations by
using the phase of { $\hat{\bf h}_k(0)$} as a reference to remove
the ambiguity and for time-varying channels we assume ideal phase
tracking. Alternatively, differential modulation can be used to
account for the phase rotations.  {The tuning parameter $\alpha$,
the forgetting factor $\beta$ and the weighting parameter $\tau$
required by the time-varying bounds described in Section IV have
been obtained by experimentation and chosen such that the
performance of the algorithms is optimized. The update rate (UR) has
been computed by counting, for each simulation trial $t$, the number
of updates ($N_{u,t}$) performed and then dividing it by the number
of received symbols ($N_{s,t}$). Then, the $UR$ is given by $UR =
\frac{1}{T}\sum_{t}^{T} \frac{N_{u,t}}{N_{s,t}}$, where $T=1000$ is
the total number of trials.}

\subsection{MSE Analytical Results}

The aim of this part is to verify the validity of the analytical
results obtained in Section V. Specifically, we shall evaluate the
analytical MSE obtained with the analytical formulas in
(\ref{emse_ss}) and (\ref{emse_tr}), and compare them to the results
obtained by simulations. We consider first a scenario with fixed
channels ($f_d=0$) and assess the MSE curves using (\ref{emse_ss})
against the number of received symbols and also versus the
signal-to-noise ratio (SNR) defined by $E_b/N_0$, as shown in Fig.
\ref{figmse1}. The results show that the analytical curves match
very well those obtained via simulations, showing the validity of
our analysis and assumptions.

\begin{figure}[!htb]
\begin{center}
\def\epsfsize#1#2{1\columnwidth}
\epsfbox{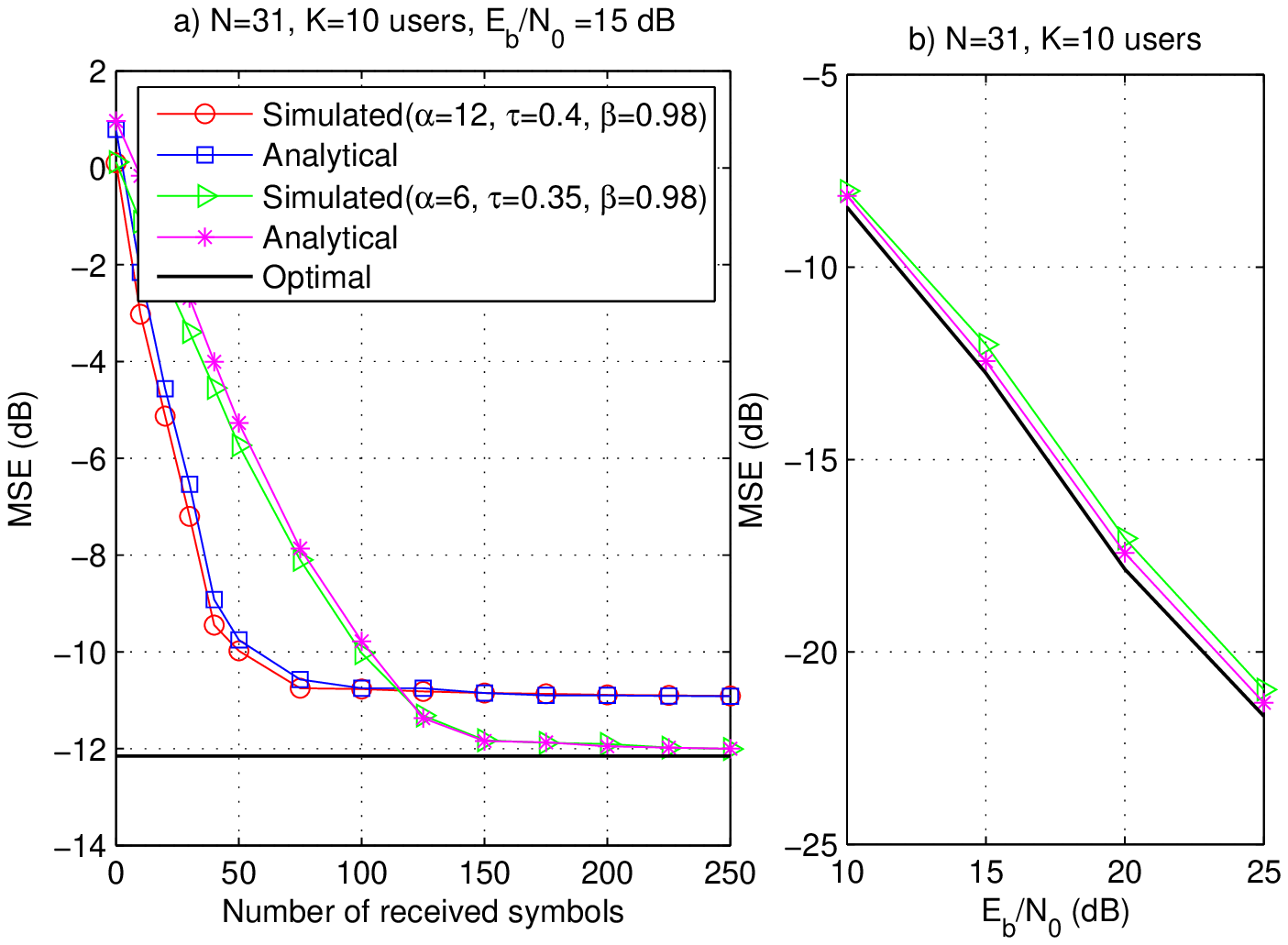} \caption{MSE performance against a) the number of
received symbols at $E_{b}/N_{0}=15 dB$ and b)  $E_b/N_0$ (dB).}
\label{figmse1}
\end{center}
\end{figure}

The tracking analysis of the proposed SM-CCM-SG algorithm in a
time-varying multipath fading channel is discussed next. We consider
the same scenario as before except for the fact that the channel is
now time-varying and the analytical results of the tracking analysis
in (\ref{emse_tr}) are employed. We consider a Jakes' model with a
typical normalized fading rate $f_dT$. We compute ${\rm
Tr}({\boldsymbol Q})$ with the aid of $J_0 (2 \pi f_d T)$
\cite{rappa}, which is the autocorrelation function of the Jakes'
model and where $J_0$ is the zero-order Bessel function of the first
kind.  {A comparison of the curves obtained by simulations and by
the analytical formulas is shown in Fig. \ref{figmse2}. Similarly to
the case of fixed channels, the analytical and simulated curves
agree and show the validity of the proposed formulas. Nevertheless,
the curves indicate a higher misadjustment as compared to the curves
in Fig. \ref{figmse1} due to the time-varying process and extra
effort of the adaptive filters to track the channel variations.}

\begin{figure}[!htb]
\begin{center}
\def\epsfsize#1#2{1\columnwidth}
\epsfbox{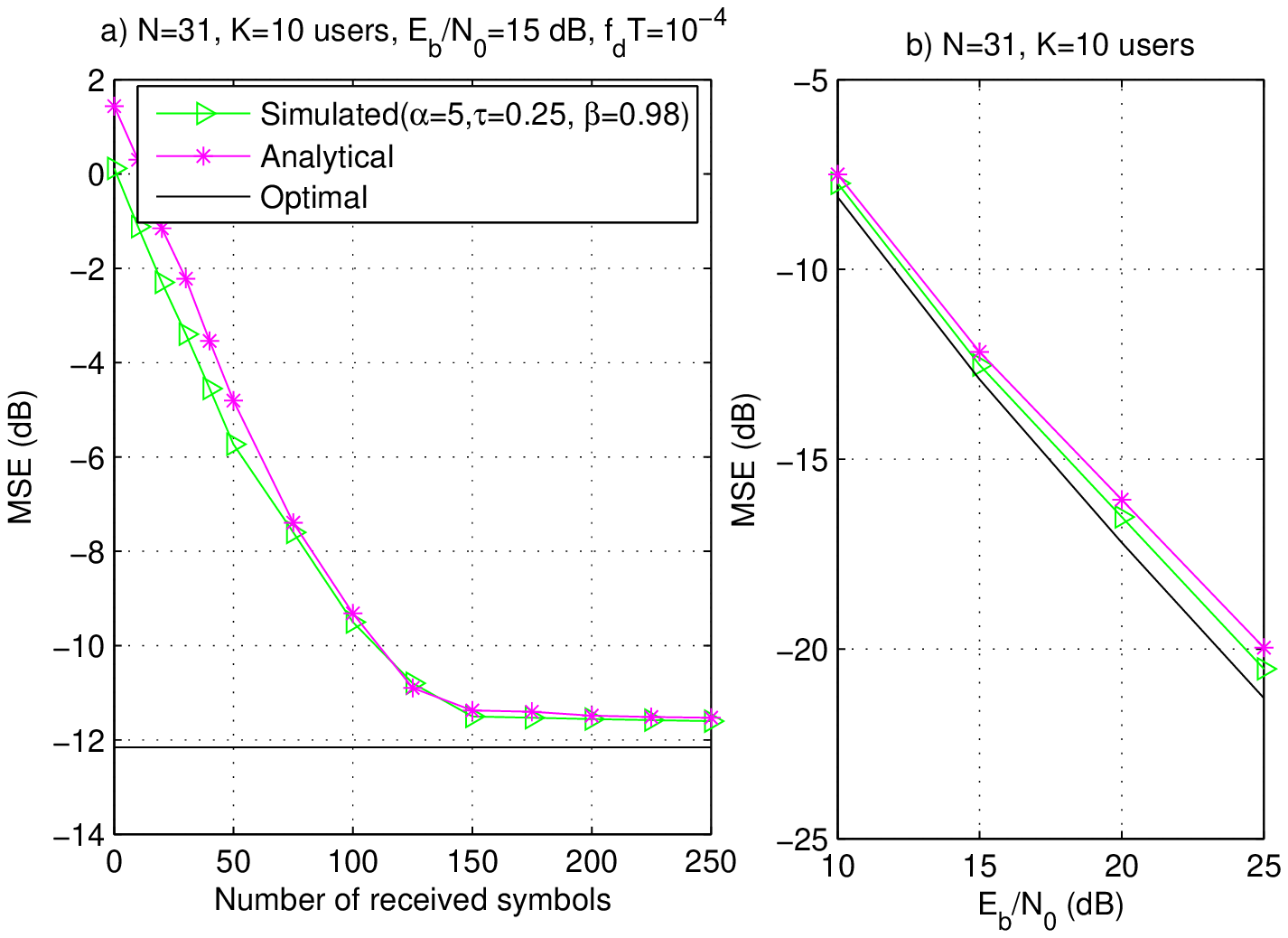} \caption{MSE performance against a) the number of
received symbols at $E_{b}/N_{0}=15 dB$ and b) $E_b/N_0$ (dB).}
\label{figmse2}
\end{center}
\end{figure}

\subsection{Interference Power and Channel Estimation}

\begin{figure}[!htb]
\begin{center}
\def\epsfsize#1#2{1\columnwidth}
\epsfbox{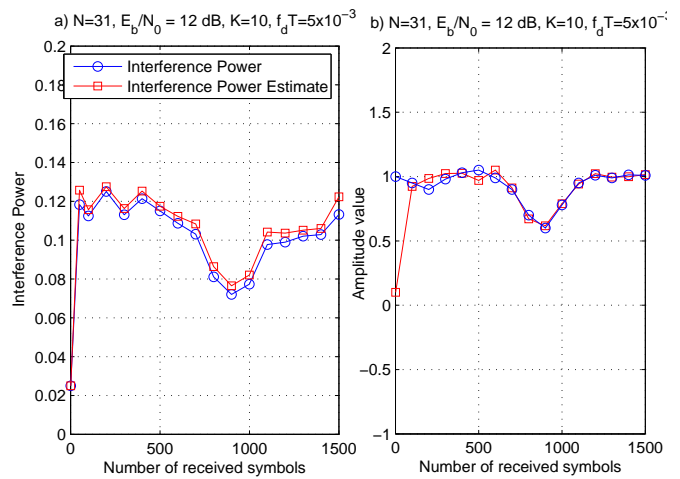} \caption{Performance of the a) interference power
estimation and tracking and b) amplitude estimation algorithms at
$E_{b}/N_{0}=12 dB$.} \label{int_pow}
\end{center}
\end{figure}

 {At this point, we wish to evaluate the effectiveness
of the proposed algorithms for estimating and tracking the
interference power and the amplitude}. To this end, we have carried
out an experiment, depicted in Fig.  {\ref{int_pow}}, where the
proposed algorithm estimates of the MAI and ISI powers and the
amplitude of the desired user are compared to the actual
interference power and amplitude. The results show that the proposed
algorithms are very effective for estimating and tracking the
interference power in dynamic environments, as depicted in Fig.
 {\ref{int_pow}}.

{We assess the proposed channel estimation (CE) algorithm, called
SM-CCM-CE, with the time-varying bounds PDB and PIDB, and also with
a fixed bound, and compare them to the CMV-based method (CMV-CE)
reported in \cite{douko} and the subspace algorithm of
\cite{wang&poor} in terms of the MSE between the actual and the
estimated channels using the following dynamic scenario. The system
has initially $10$ users, the power distribution among the
interferers follows a log-normal distribution with associated
standard deviation of $3$ dB. After $1500$ symbols, $6$ additional
users enter the system and the power distribution among interferers
is loosen to $6$ dB. The results, shown in Fig \ref{ce}, reveal that
the proposed SM-CCM-CE algorithm outperforms the CMV-CE technique
reported in \cite{douko} because SM-CCM-CE uses a variable
forgetting factor. The proposed SM-CCM is also computationally
simpler than CMV-CE due to the sparse updates, whereas it is
substantially less complex than the subspace method of
\cite{wang&poor}. This is because the SM-CCM-CE needs to compute the
principal eigenvector of the $L_p \times L_p$ matrix ${\boldsymbol
C}_k \hat{\boldsymbol R}_k^{-1}[i]{\boldsymbol C}_k$, while the
subspace technique in \cite{wang&poor} requires the principal
eigenvector of the $M \times M$ matrix $\hat{\boldsymbol
R}^{-1}[i]$. Since $L_p \ll M$ in most practical scenarios SM-CCM-CE
is typically considerably simpler than the subspace method.}

\begin{figure}[!htb]
\begin{center}
\def\epsfsize#1#2{1\columnwidth}
\epsfbox{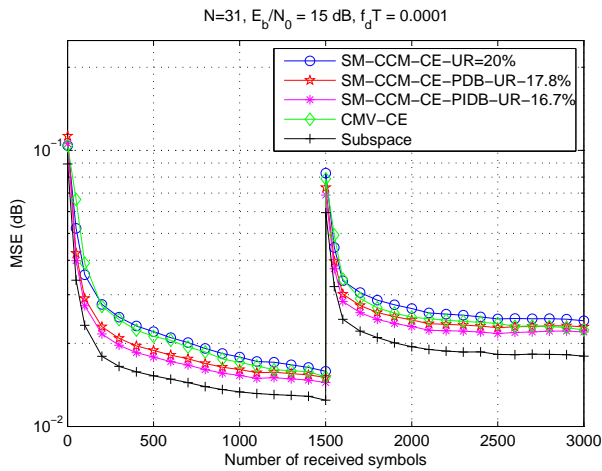} \caption{MSE performance of channel estimation
versus number of received symbols in a dynamic scenario where
receivers operate at $SNR=E_{b}/N_{0}=15$ dB for the desired user.}
\label{ce}
\end{center}
\end{figure}

\subsection{BER Performance}

The SM-CCM algorithms are assessed in a non-stationary environment
where users enter and exit the system, as depicted in Figs.
\ref{conv_sg} and \ref{conv_rls}. The system starts with $2$
interferers with $7$ dB above the desired user's power level and $5$
interferers with the same power level of the desired one, which
corresponds to the signal-to-noise ratio $E_{b}/N_{0}=15$ dB. At
$1000$ symbols, $2$ interferers with $10$ dB above the desired
signal power level and $1$ with the same power level enter the
system, whereas $1$ interferer with $7$ dB above the desired signal
power level leaves it. At $2000$ symbols, $1$ interferer with $10$
dB above, and $5$ interferers with the same power level of the
desired signal exit the system, while $1$ interferer with $15$ dB
above the desired user enters the system.  {The results for $100$
runs show that the proposed SM-CCM-RLS algorithm achieves the best
performance, followed by the proposed SM-CMV-RLS recursion, the
CCM-SG and the CMV-SG methods. In summary the SM-CCM algorithms
outperform the SM-CMV techniques in all scenarios and the SM-CCM-RLS
algorithm is the best among the analyzed algorithms.}

 {In a near-far scenario the eigenvalue spread of the
covariance matrix of the received vector ${\boldsymbol r}[i]$ is
large, affecting the convergence performance of the SG algorithms
with fixed step size and making it very difficult to compute a
pre-determined value \cite{diniz} for the step size. In this case,
the SM-CCM algorithms are able to deal with near-far situations
since they adopt variable step size or variable forgetting factor
mechanisms, ensuring good tracking performance in dynamic scenarios
and an improved performance over the existing CCM-type algorithms.}
In addition, due to their data-selective update feature the SM
algorithms can save significant computational resources as they only
require parameter updates for about $20 \%$ of the time for the
SG-type recursions and around $12 \%$ for the RLS-based techniques.

\begin{figure}[!htb]
\begin{center}
\def\epsfsize#1#2{1\columnwidth}
\epsfbox{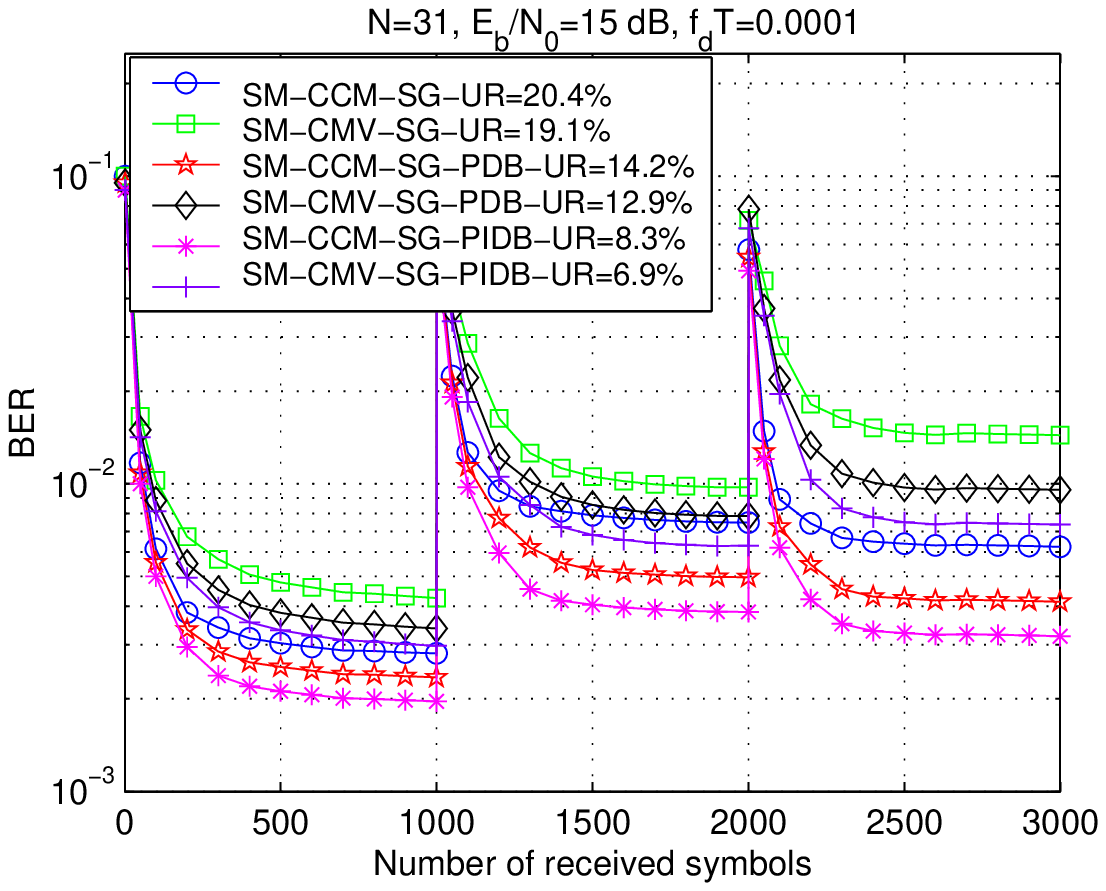} \caption{BER performance of the algorithms versus
number of symbols for a non-stationary scenario. The parameters are
$\gamma = 1.3$ for the SM-CMV, $\gamma=0.65$ for the SM-CCM.}
\label{conv_sg}
\end{center}
\end{figure}

\begin{figure}[!htb]
\begin{center}
\def\epsfsize#1#2{1\columnwidth}
\epsfbox{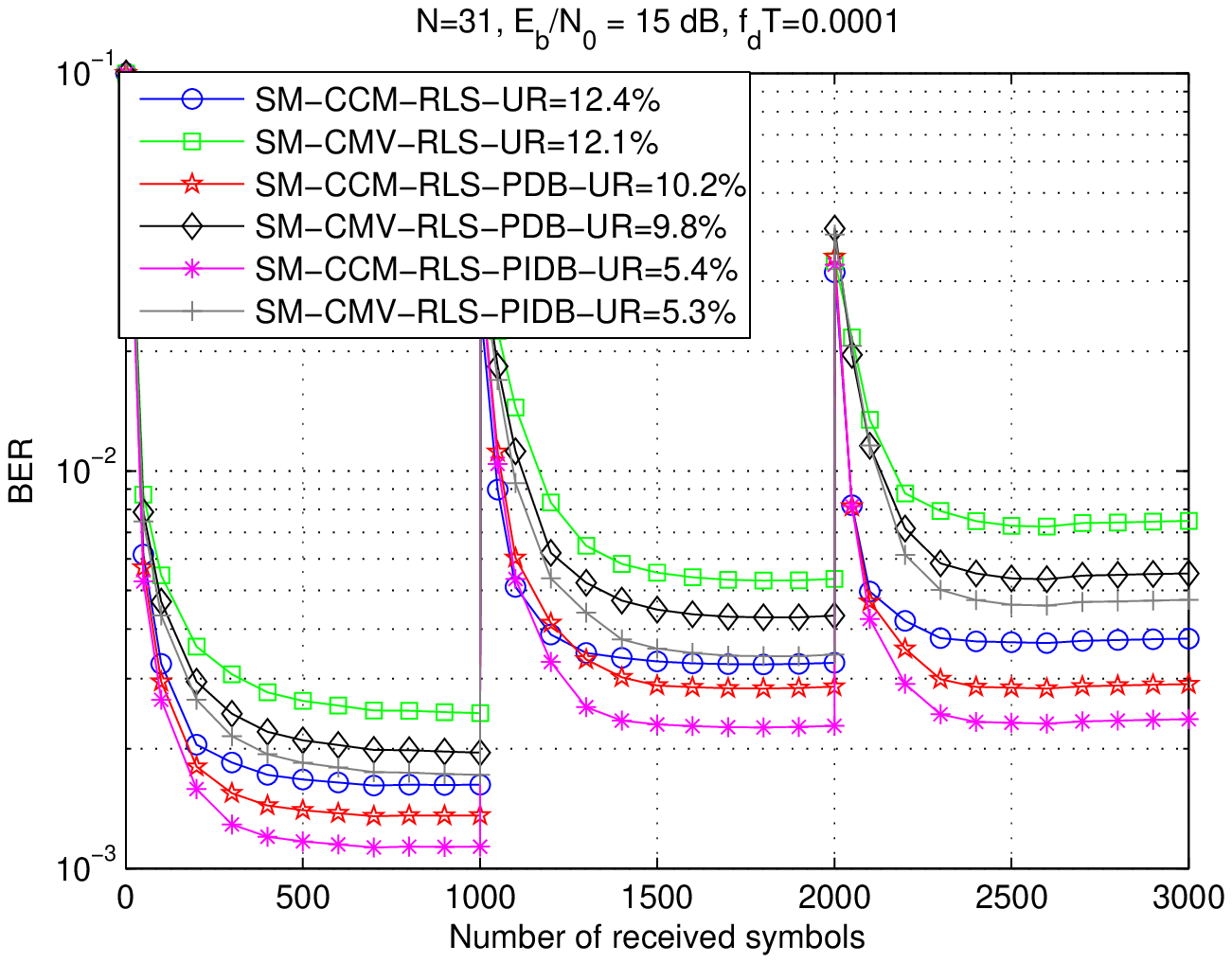} \caption{BER performance of the algorithms versus
number of symbols for a non-stationary scenario. The parameters are
$\gamma = 1.3$ for the SM-CMV, $\gamma=0.65$ for the SM-CCM.}
\label{conv_rls}
\end{center}
\end{figure}

 {The same scenario illustrated in Fig. \ref{conv_sg}
is considered for the SM-CCM algorithms with time-varying error
bounds, as shown in Figs. \ref{tv_sg} and \ref{tv_rls}. The results
indicate that the proposed time-varying bounds are capable of
improving the performance of SM algorithms, while further reducing
the number of updates. The RLS-type algorithms outperform the
SG-based techniques as expected and have lower $UR$. Moreover, the
algorithms with the PIDB approach achieve the best performance,
followed by the algorithms with the PDB method and the SM recursions
with fixed bounds. With respect to the $UR$, the PIDB technique
results in the smallest number of updates, followed by the PDB
approach and the fixed bound.}

\begin{figure}[!htb]
\begin{center}
\def\epsfsize#1#2{1\columnwidth}
\epsfbox{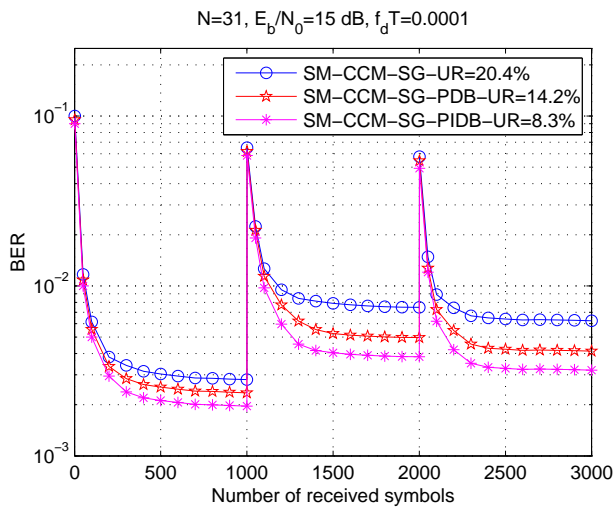} \caption{BER performance of the SG algorithms
versus number of symbols for a non-stationary scenario with
time-varying bounds. The parameters are $\alpha = 8$, $\tau=0.35$
and $\beta=0.95$ for the time-varying bounds.} \label{tv_sg}
\end{center}
\end{figure}

\begin{figure}[!htb]
\begin{center}
\def\epsfsize#1#2{1\columnwidth}
\epsfbox{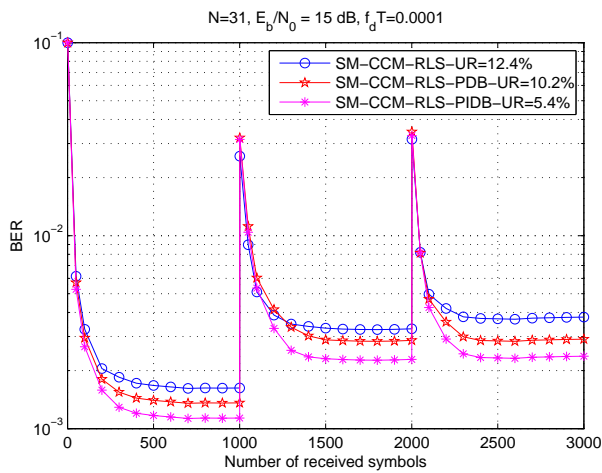} \caption{BER performance of the RLS algorithms
versus number of symbols for a non-stationary scenario with
time-varying bounds. The parameters are $\alpha = 8$, $\tau=0.35$
and $\beta=0.95$ for the time-varying bounds.} \label{tv_rls}
\end{center}
\end{figure}

The BER performance versus $E_{b}/N_{0}$ and number of users is
shown in Fig. \ref{berxsnr}.  {We consider data packets of $P=1500$
symbols and compare the proposed SM-CCM-RLS algorithm with a fixed
bound, with the PIDB mechanism, the training-based BEACON algorithm
with the PIDB mechanism \cite{delamaresmf} and the linear MMSE
receiver that assumes perfect knowledge of the channels and the
noise variance. We have measured the BER after $200$ independent
transmissions and the BEACON algorithm
\cite{nagaraj},\cite{delamaresmf} is trained with $200$ symbols and
is then switched to decision-directed mode. The curves illustrate
that the proposed blind SM-CCM-RLS algorithm can approach the
performance of the supervised BEACON algorithm and that of the
linear MMSE receiver.}

\begin{figure}[!htb]
\begin{center}
\def\epsfsize#1#2{1\columnwidth}
\epsfbox{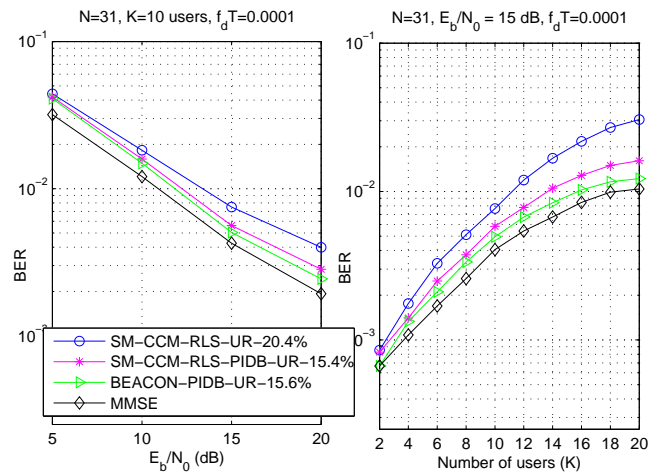} \caption{BER performance versus (a)
$E_{b}/N_{0}$ with $K=10$ users and (b) number of users (K) at
$E_{b}/N_{0}=15$ dB .} \label{berxsnr}
\end{center}
\end{figure}

\section{Conclusions}

 {We have proposed SM-CCM adaptive algorithms for
interference suppression in DS-CDMA systems. We have analyzed the
optimization problem that gives rise to the SM-CCM algorithms and
devised analytical expressions to predict their MSE performance with
a good accuracy. The proposed SM-CCM algorithms are equipped with
variable step sizes and forgetting factors and are only updated for
a small fraction of time without incurring any performance
degradation. A blind framework for SM estimation that takes into
account parameter estimation dependency and MAI and ISI for
multiuser communications has also been introduced to provide a
time-varying bound that is robust against channel and SNR
variations. Simulations have shown that the proposed SM-CCM
algorithms outperform previously reported blind techniques for
several scenarios.}

\section*{Acknowledgement}

The second author is grateful for the financial support provided by
CNPq, a Brazilian research council, and FAPERJ, a Rio de Janeiro
state research funding agency.

\end{document}